*Submitted to Journal of Phase Equilibria and Diffusion*

# On Gibbs Equilibrium and Hillert Nonequilibrium Thermodynamics

Zi-Kui Liu

Department of Materials Science and Engineering, The Pennsylvania State University,

University Park, Pennsylvania 16802, USA

**Abstract:**

During his time at Royal Institute of Technology (Kungliga Tekniska högskolan, KTH) in Sweden, the present author learned nonequilibrium thermodynamics from Mats Hillert. The key concepts are the separation of internal and external variables of a system and the definitions of potentials and molar quantities. In equilibrium thermodynamics derived by Gibbs, the internal variables are not independent and can be fully evaluated from given external variables. While irreversible thermodynamics led by Onsager focuses on internal variables though often mixed with external variables. Hillert integrated them together by first emphasizing their differences and then examining their connections. His philosophy was reflected by the title of his book "Phase Equilibria, Phase Diagrams and Phase Transformations" that puts equilibrium, nonequilibrium, and internal processes on equal footing. In the present paper honoring Hillert, the present author reflects his experiences with Hillert and his work in last 40 years and expresses his gratitude for all the wisdom and support from him in terms of "Hillert nonequilibrium thermodynamics" and discusses some recent topics that the present author has been working on.

## Table of Contents

## 1 Introduction

The present author learned thermodynamics twice before joining Royal Institute of Technology (Kungliga Tekniska Högskolan, KTH, Sweden) in 1987. The first time was at the Central South University (CSU, formerly Central South Institute of Mining and Metallurgy, Changsha, China) as an undergraduate student, and the second time at the University of Science and Technology Beijing (USTB, formerly Beijing Steel and Iron Institute, Beijing, China) as a graduate student in its Master program. It was after third time at KTH, the present author learned from Hillert that all the formalisms taught before were *only* for equilibrium thermodynamics though commonly applied to nonequilibrium processes such as diffusion and phase transformations, which inevitably causes many confusions. It took several more decades of teaching and research, particularly after joining The Pennsylvania State University (PSU) in 1999, for the present author to integrate Hillert's nonequilibrium thermodynamics [1,2] with quantum and statistical mechanics in terms of the zentropy theory to account for the multiscale total entropy in a system [3,4] and apply it to address irreversible thermodynamics in terms of fluxes and driving forces [5–7].

In addition to great science learned from Hillert over the years, the present author and his family also greatly benefited from the diverse care that Hillert and his family offered for the decade when they worked at KTH. In the present overview paper honoring Hillert, the present author reflects his experiences and interactions with Hillert and presents his views on the enormous impact that Hillert had on material science in general and thermodynamics in particular with the term "Hillert nonequilibrium thermodynamics" or simply "Hillert thermodynamics" recently proposed by the present author [7].

## 2 Present author's early career path along thermodynamics

To show the impact of Hillert thermodynamics on the work of the present author, it is necessary to give a short description of present author's early experiences in thermodynamics. While taking the thermodynamic class (called physical chemistry) at CSU, the deep voice of the instructor along with complex solution thermodynamics lulled students to sleep in those very hot days in Changsha. While the first mid-term exam of the class was very disappointing but excited the present author to master those abstract concepts by studying the subject with several close friends in the class. This turned out to be a critical factor in determining the career path at the end of present author's junior year at CSU.

While in middle school, the present author's physics teacher, Mr. Manzhao Yang, led a group of students to build a chemistry lab from scratch as there was no such lab after the culture revolution in the remote mountain area in China. That was the first time that physics was connected with chemistry for the present author. So, the present author wanted to study chemistry at CSU, but was assigned to the major of pelletization of iron ore fines as feed to blast furnace or direct reduction systems to produce iron. During the junior year at CSU, the present author and his friends started to discuss what to do after graduation and decided to go to graduate school. However, there was only one opening for the graduate study in our major of 60 classmates. It was evident that many needed to change majors for graduate studies.

The present author's decision to change the major to materials science and engineering was first based on physical chemistry being the foundational course for both majors. The decision was further solidified by a class on "Heat Treatment of Metals" and associated labs on metalworking practices. The enthusiasm of the teacher and her description of challenges and debates on bainite transformations inspired the present author to look for research programs on phase transformations. The ultimate choice was USTB where there were three professors with their research activities on phase transformations in metals. The admission to the USTB graduate program was only made possible with the score of 94 out of 100 on the entrance exam of physical chemistry by solving a challenging problem on vapor, partial, and total pressures, which also showed up many years later at the present author's research at PSU [8–13] and is discussed in Section 8.2.

It turned out that present author's Master thesis [14] was on experimental investigations on phase transformations of bainite in high strength low alloy steels including research activities on heat treatment, rolling, etching of austenite grain boundary, transmission electron microscopy (TEM), mechanics testing, and computer programming using punch cards and IBM PC without hard disk to simulate the orientation relationships between bainite and austenite. The present author is grateful for his advisor, Wenxuan Cui, for giving him the freedom and financial support to explore those new capabilities, which did sometimes result in costly repair of furnace, cutting into a finger by electric discharge machine in early hours after midnight, operating TEM overnight at the TEM center of Chinese Academy of Science to catch the deadlines, or pile of computer printing out on orientation relationships. During those research activities, the present author was wondering about the interface conditions between bainite and austenite and later learned the concept of local equilibrium discussed in Hillert's lectures given at USTB in 1980 with its Chinese translation published in 1984 [15], which turned out to be the core component of the present author's PhD study at KTH in terms of both measurements of local compositions at phase interfaces by TEM that confirmed the local equilibrium between ferrite and cementite, and theoretical and computational simulations [16–20].

## 3 Coming to and Experiencing KTH

The present author's wife, Weiming Huang, who was the smartest person in the graduate class at USTB with perfect scores in all courses, was recommended for a scholarship to study for her PhD in Hillert's group in 1986. Subsequently, Hillert wrote a personal letter to the present author's department head, Guoliang Chen, to request him to let the present author to go to KTH for his PhD study under the supervision of John Ågren on phase transformations. After seven days of train and ship travel from Beijing through Moscow and Helsinki, the present author arrived at Stockholm in the beautiful summer of 1987 for his PhD study at KTH. Even more heartwarming was that Hillert and his wife Gerd let us use their lovely house on the Lidingö island overlooking the Saltsjön that is connected to the Baltic Sea with detailed instructions on how to use dish washer, washing machine and other facilities in the house when they were on summer vacation in Gothenburg, Sweden.

At every Christmas, Hillert invited students, visiting scholars, and their families to his house to celebrate the holiday with his wife and his family of four children and the gradually increasing number of grandchildren. The festivity was full of fun memories including traditional Swedish Christmas food, e.g., surströmming and meat ball with lingonberry, and drink e.g., julmust, dancing around in the house, singing Swedish folk songs, and playing various games. Two photos of the 1987 Christmas party are shown in Figure 1.

*Figure 1: Christmas photos at Hillert's house in December 1987, (a) dancing around in the house, and (b) singing in the kitchen while someone was hiding an item for others to find.*

The excellent equality and equity in Sweden are well known in the world. This was also reflected in the friendly environment established by Hillert in the department. Everyone was called by his/her first name, including Mats for Hillert and John for Ågren. Hillert's office door was usually open, and students could walk in asking questions or for a quick discussion. Except Hillert, most people went to canteens on campus for lunch together, often enjoying a small bottle of beer there. The twice daily coffee breaks and weekly group seminar with cakes were regarded as truly civilized by our guests from the US and were great for social chatting and scientific discussion that brought people close to each other. The first group photo including the present author was take in 1988 as shown in Figure 2 along with the one in 1992 when the present author graduated. The last photo with Hillert was a group photo with Ågren and the present author's students at the CALPHAD conference in Lidingö where Hillert's house is located (see Figure 3). Hillert participated at the whole CALPHAD conference from Sunday to Friday.

*Figure 2: Group photos in (a) 1988 and (b) 1992, KTH, Sweden.*

*Figure 3: Group photo on May 23, 2022 at the CALPHAD conference in Lidingö, Sweden.*

Furthermore, everyone in the department had similar number of annual vacation days and enjoyed long summer away from office including the important midsummer celebration. The department celebrated many events, particularly the PhD graduations. In addition to the PhD graduation ceremony and banquet every other year presented by KTH in the city hall where the Nobel banquet is held, a big celebration was either organized at the PhD awardee's home or in a restaurant with everyone in the department and awardee's family members invited. To celebrate the department's scientific achievements, the list of publications was hanged on the door of the mail cabinet in the corridor outside Hillert's office, which was around 300 papers in 1987. The celebrations of the PhD graduations of the present author's wife and the present author and the births of the two sons in Sweden were some of the truly loving memories with some photos shown in Figure 4

*Figure 4: (a) Banquet after Weiming Huang's PhD defense on October 30, 1990, (b) & (c) Banquet after the present author's PhD defense on May 21, 1992, (d) Dinner on the island of Hillert's house during the visit to Sweden on June 17, 2013.*

The present author learned a great deal from Hillert how to communicate with people. Before discussing criticisms on a piece of work, Hillert often started with "I must be very frank with you". Hillert deeply cared about his students and others around him. He mentioned that while educating students one must teach them how to swim and at the same time ensure that they do not drown. It was several years after the graduation, the present author was told that even though Hillert was not directly involved with the present author's research, that Hillert often discussed with Ågren the present author's research directions, challenges that the present author was facing, and how to further the present author's research progresses [21]. Following Hillert's approaches, the present author often tells his students that "I will be very frank and open to you by asking direct questions and giving you my candid comments and suggestions. After you graduate, you will be an independent researcher and will no longer hear unsolicited advice from me". The present author has developed his TKC theory based on those approaches [22] with the motto being "Do Better Than Our Best" [23], where TKC represents Thermodynamics for evaluating driving force, Kinetics for understanding barriers, and Crystallography for managing coherency with one's surroundings.

Another important research approach that Hillert instituted was to use the Science Citation Index (SCI), which is now Web of Science. While one often checks the references cited in an interesting paper to understand the history of the topic, Hillert emphasized that it is even more important to check the newer publications that cited this interesting paper in SCI. He pointed out that it is likely that the paper is also interesting to others if it is interesting to you and thus cited by others in their publications. It is therefore crucial to read those newer publications to understand the progress of the topic, avoid repeating what others have accomplished, identify what is missing, and make better and informed decisions on what to do next. It is like standing on the shoulders of giants as Hillert often mentioned. At that time, SCI was on hard copies. It was a lot of fun flipping through those thick books in the KTH library to find the papers which cited an interesting paper and understand how scientists think and dissect a problem, develop approaches to address the problem, and improve understanding of the problem. The integration of forward and backward approaches has been guiding the present author's research ever since, including the funded projects [24–26] and the development of the TKC theory [22].

## 4 Teaching of kinetics and thermodynamics by Hillert

The two courses that Hillert taught at KTH fundamentally changed the present author's views on thermodynamics and phase transformations: "Diffusion Theory in Alloys" for undergraduate students and "Thermodynamics" for graduate students. Hillert delivered both courses in English, which greatly helped the non-Swedish speaking students to understand the complex concepts. Though the contents of the undergraduate course were similar to those in the Chinese translation [15], Hillert's lectures were much more detailed and made the concepts much more vivid, particularly the flux equations for diffusion, the local equilibrium at a phase interface, and the concentration profile in front of a moving interface, which was termed as the bow wave in front of a moving ship by Heyi Lai at USTB who visited KTH before and translated Hillert's lecture notes [15].

A striking knowledge acquired during the diffusion course was the morphology of pearlite. As discussed by De Graef, Kral, and Hillert in 2006 [27], Hillert in 1950s showed that pearlitic ferrite can have the same crystalline orientation as adjacent proeutectoid ferrite in hypoeutectoid steels, while in hypereutectoid steels, pearlitic cementite can have the same orientation as adjacent proeutectoid cementite. Hillert proposed that pearlite forms by the cooperative growth of ferrite and cementite with incoherent interfaces with the austenite matrix. Through sectioning 242 times a pearlite colony in a carburized electrolytic iron and combining the micrographs into a movie, it was observed that ferrite and cementite cooperate and grow together into the austenite as lamellar pearlite through branching and without new nucleation. Hillert believed that this is probably similar in any other eutectic or eutectoid transformation. Hillert further pointed out that ferrite and cementite in pearlite usually have random orientation relationship with the parent austenite, while ferrite in bainite has a coherent interface with austenite probably due to the strain energy and sluggish diffusion at low temperature.

On the other hand, this issue was not clear for a long time for bainite where a repeated nucleation process and sympathetic nucleation were often discussed [28,29], which could be possible when branching is no longer possible [30]. During his research at USTB, the present author tried to study the morphology of bainite in three dimensions by polishing and etching two perpendicular surfaces as shown by the micrographs taken in scanning electron microscopy (SEM) in Figure 5 (a)-(c) along with various morphologies shown on micrographs from optical microscopy (OM) in Figure 5 (d) and (e). In preparing for the present paper, the present author found two manuscripts with one containing micrographs shown in Figure 5, which were heavily marked by Hillert [31], but somehow they were never published.

*Figure 5: SEM micrographs of bainite on two-perpendicular surfaces showing the three-dimensional morphologies of ferrite in bainite: (a) a single straight plate in SEM, (b) a single branched plate in SEM, (c) a package with plate shape on upper surface and branched and irregular shapes on lower surface in SEM, (d) a package with 2D plate shape in OM, and (e) a package with 2D rectangle shape. Micrographs were taken from a steel of Fe-0.15C-1.54Mn-0.53Mo-0.015N-0.021RE-0.065V-0.039Al-0.019P-0.015S-0.51Si (wt%) homogenized at 1200℃ for one hour and held at 430℃ for 5 seconds followed by quenching into water* [31].

The thermodynamics course by Hillert was truly eye-opening to the present author. During his time as a class master at USTB, the present author helped his students to understand thermodynamics better by giving extra recitations. Hillert's class made the present author realize that his understanding of thermodynamics was for equilibrium systems only. To encourage students to dive deeper into the abstract concepts, Hillert mentioned that "if you can truly master the contents in this class, you will be a few in the world who understand thermodynamics", which was truly inspiring.

While details of Hillert thermodynamics will be discussed in Section 6, there are a number of features of Hillert's teaching that the present author remembers well and benefited greatly from it. Hillert often reminded the class that the fundamental principles of thermodynamics are rather simple with only a few rules, and the complexity is in their applications to multicomponent materials. One such example is the chemical potential. Its definition is a simple partial derivative of internal or free energy with respect to the number of moles of a component. However, for a system with total one mole of components, the moles of all components are not independent. Consequently, the derivative of a molar energy of the system with respect to compositions becomes rather complicated, resulting a complex formula for chemical potential in terms of molar internal or free energy. Hillert pointed out that it is important to understand those abstract concepts through virtual or thought experiments and then link them through computer programs, which was probably why the software packages from his group were so general and applicable to many complex applications.

Hillert did vivid demonstrations in his lectures. For example, to show the concept of instability, he would stand on the lecture table and drop a piece of paper which developed instability when the falling speed became high. Another interesting observation by all his students was that everything was clearly understood during his lectures, but it was hard to figure where to start on homework assignments. Hillert said that this was to develop critical thinking skills to use learned knowledge to solve new problems that one has not seen before. For classes with a small number of students, Hillert engaged students by asking questions individually and encouraging everyone to participate.

While waiting for various facilities becoming available for sample preparation and composition measurements in TEM, the present author spent some time to understand a number of interesting concepts lectured by Hillert in the class through intensive discussions with Hillert and Ågren, particularly paraequilibrium, coherent equilibrium, and Le Chatelier's principle, which resulted in several publications [17,32–35]. The discussions with Hillert on Le Chatelier's principle and its application to the controversial ammonia reaction were particularly inspiring in that Hillert demonstrated that the principle is valid for infinitesimal changes only [36] which was incorporated into the paper submitted by the present author and resulted in the only joint publication that the present author had with Hillert [35]. The discussions on Gibbs phase rule made the present author understand the importance to define molar quantities and potentials. The investigations and publications on Le Chatelier's principle and Gibbs phase rule would not have been completed without Hillert's guidance and those "very frank" criticisms, from which the present author benefited immensely.

Another interesting topic inspired by Hillert's class was the Maxwell relation. While it was easy to understand the relation through the second derivatives of internal or free energy, Hillert mentioned that its practical usefulness had not been demonstrated. This remained in the present author's mind ever since until very recently that the present author realized that those derivatives represent experimentally observed cross phenomena with one being relatively easy to measure experimentally and the other being ideal for theoretically investigations as discussed in Section 7.2 [5].

Hillert revised his lecture compendium every time he taught it, and his secretary Brita Gibson would type it every time using her typewriter. Gibson was always elegant with a touch of nobility as shown in Figure 2, and the sound from her typing sounded like music and could be heard when one walked through the corridor. Hillert and Gibson knew each other before the KTH time. At Gibson's retirement party, Hillert showed a great sense of humor. He told the story that Gibson often typed "It is worth nothing" at the places where it should be "It is worth noting" and joked that Gibson might be right after all which made everybody laugh. After retirement in 1991, Hillert started to learn using computers and published the first edition of his book titled "Phase Equilibria, Phase Diagrams and Phase Transformations" in 1998 and had been to office regularly until he passed away on November 2, 2022. It will become evident that this title reflects the origin of Hillert's significant contribution to thermodynamics in terms of the title of the present paper. One interesting feature of the first edition of this book is that equations are not numbered, and we joked that Hillert wanted us to remember all the equations and know where they are in the book. The present author used this book and its second edition in 2007, in which all equations are numbered, as the textbook for the graduate thermodynamics course at PSU until the class was taken over by another faculty during his sabbatical leave in 2014 and 2015 and was unfortunately not given back to him to teach afterwards. Nevertheless, the present author published his own textbook titled "Computational Thermodynamics of Materials" in 2016 with the chapter on first-principles calculations based on the density functional theory (DFT) contributed by co-author Yi Wang [37].

## 5 Gibbs equilibrium thermodynamics

One of the foundational contributions that Gibbs made to thermodynamics was by replacing the heat change in the first law of thermodynamics by the entropy change using the relation defined by Clausius for an reversible process [38]. By avoiding the issue of inequality in the second law of thermodynamics for nonequilibrium systems with internal processes, Gibbs was able to obtain the combined law of thermodynamics for a closed *equilibrium* system as follows

$$dU = dQ + dW = TdS - PdV \qquad \textit{Eq. 1}$$

where $dU$ is the internal energy change of the system, $dQ$, $dW$, $dS$ and $dV$ are the increases of heat, work, entropy, and volume of the system controlled from the surroundings, and $T$ and $P$ are the temperature and pressure. Gibbs [39–41] termed Eq. 1 as the fundamental equation and emphasized that the relation between U, S, and V carries more knowledge than the other combinations of quantities in Eq. 1 such as V, P and T proposed by Thomson earlier. Its importance was greatly appreciated by Maxwell who hand-made two three-dimensional models to represent the function of $U(S,V)$ with one copy sent to Gibbs [42] and one kept in Cavendish Laboratory at the University of Cambridge [6].

For an open multicomponent *equilibrium* system under hydrostatic pressure, Gibbs [40,41] introduced the concept of chemical potential for each component and presented Eq. 1 as follows

$$dU = TdS - PdV + \sum_{i=1}^{c} \mu_i dN_i = \sum_{a=1}^{c+2} Y^a dX^a \qquad \textit{Eq. 2}$$

where $\mu_i$ is the chemical potential of component $i$, $N_i$ is the moles of component $i$, and $Y^a$ and $X^a$ represent the pairs of conjugate variables with $Y^a$ for potentials, such as $T$, $P$, and $\mu_i$, and $X^a$ for molar quantities, such as $S$, $V$, and $N_i$ [2,6,37]. Eq. 2 shows that $X^a$ are independent variables of the internal energy and are termed as its natural variables because they emerge naturally through the combination of the first and second laws of thermodynamics for equilibrium systems [1,2]. It is noted that Gibbs called $\mu_i$ and $P$ potentials, but not $T$ probably due to its obviousness. Gibbs further showed that every potential is homogeneous in a system under equilibrium and defined today's free energies such as Helmholtz energy, enthalpy, Gibbs energy, Gibbs-Duhem equation, Gibbs phase rule, Clausius-Clapeyron equation, and stability of an equilibrium.

Most thermodynamic textbooks are based on Eq. 1 and Eq. 2, which are used to further derive various properties and applications, sometimes erroneously for nonequilibrium systems as pointed out by the present author [5–7]. The first portion of Eq. 1 does not require that the system is under equilibrium and concerns only the exchange of heat and work between the system and its surroundings. It does not concern what happens inside the system, i.e., the internal processes. This was probably why irreversible thermodynamics had to be developed separately though it was unfortunately based on phenomenological observations as pointed out by Hillert [1,2] and less fundamental than the first and second laws of thermodynamics as mentioned by Onsager [43] and commented by Balluffi et al. [44].

## 6 Hillert nonequilibrium thermodynamics

Hillert majored in chemical engineering and worked at the Swedish Institute for Metals Research before coming to the US for his ScD at MIT. In his thesis [45,46], Hillert developed a solid-solution thermodynamic model based on the nearest-neighbor interactions to predict the existence of a periodically modulated structure in ordering and precipitation systems, calculate the diffuse grain and domain boundaries, and experimentally investigated the spinodal transition in the Cu-Ni-Fe system. It included a kinetic treatment to calculate the time evolution of the structures in a thermodynamically unstable system.

In his book [1,2], Hillert started by defining external variables with their values that can be changed through interactions between the system and its surroundings and internal variables that change inside the system due to internal processes before the system reaches its equilibrium. Hillert then introduced the first and second laws of thermodynamics and entropy change for an open system as follows

$$dU = dQ - PdV + H_m dN \qquad \text{Eq. 3}$$

$$dS = \frac{dQ}{T} + S_m dN + d_{ip}S \qquad \text{Eq. 4}$$

$$d_{ip}S = \frac{1}{T} D d\xi \geq 0 \qquad \text{Eq. 5}$$

where $H_m$ and $S_m$ are the molar enthalpy and molar entropy of the added materials $dN$, and $d_{ip}S$ is the entropy production due to the internal process $d\xi$ with its driving force being $D$ and internal variable being $\xi$. The second law is represented by Eq. 5. It is important to note that $dS$ now contains both external and internal contributions with the first two terms and the last term in Eq. 4 for the former and the latter, respectively.

Hillert obtained the combined law of thermodynamics by combining Eq. 3 and Eq. 4 as follows

$$dU = TdS - PdV + G_m dN - Dd\xi \qquad \text{Eq. 6}$$

where $G_m$ is the molar Gibbs energy of the added materials. For a system with variable composition, Hillert introduced chemical potential as Gibbs did and obtained the following combined law of thermodynamics

$$dU = TdS - PdV + \sum_{i=1}^{c} \mu_i dN_i - Dd\xi = \sum_{a=1}^{c+2} Y^a dX^a - Dd\xi \qquad \text{Eq. 7}$$

Eq. 7 reduces to Eq. 2 when $\mathrm{D}d\xi = 0$. In a recent publication [7], the present author suggested to call Eq. 2 *Gibbs equilibrium thermodynamics* and Eq. 7 *Hillert nonequilibrium thermodynamics* because Eq. 7 includes both first and second law of thermodynamics, while Eq. 2 does not contain the information from the second law of thermodynamics and is only applicable to the special case with $\mathrm{D}d\xi = 0$, i.e., at equilibrium. They are termed as *Gibbs thermodynamics* and *Hillert thermodynamics* in the rest of the present paper for brevity.

Hillert further discussed two types of equilibria with either $D = 0$ or $d\xi = 0$ and termed the former as smooth equilibrium and the latter as freezing-in conditions where the internal variable $\xi$ becomes an independent variable of the system. The derivative of internal or free energy to $\xi$ thus gives the driving force for the change of $\xi$ as an internal process with $dX^a = 0$. Consequently, all formulations in Gibbs thermodynamics can be directly used in Hillert thermodynamics by treating $\xi$ as an independent variable in analogy to the external variables, i.e., $X^a$ in Eq. 7.

Based on Hillert's thesis [45,46], Cahn [47] systemized the theory of spinodal decomposition that enabled today's phase-field simulations [48]. Hillert further developed the theory for normal and abnormal grain growth [49], regular-solution model [50], the theory for growth during discontinuous precipitation [51], and the magnetic model for solution phases [52]. After the first CALPHAD conference at Larry Kaufman's home in Boston in 1973 (see photos in Figure 6), Hillert recruited three graduate students in 1974 to develop the Thermo-Calc and DICTRA software packages [53,54]. Since Thermo-Calc and DICTRA were based on fundamental principles of thermodynamics and diffusion with the internal variables defined in Hillert thermodynamics along with the widely used compound energy formalism [55,56], their applications have been very versatile and impactful. The present author used the PARROT module developed by Jansson [57] to perform the equilibrium calculations under various freezing-in conditions such as the transition from local equilibrium to paraequilibrium [17], coherent equilibrium [32,33], and solute drag [17,20,58–61]. All of them involve internal variables and derivatives of Gibbs energy with respect compositions that could be easily obtained from the PARROT module in Thermo-Calc. This important functionality is realized by a separate command, "calculate", in open-source PyCalphad software developed by Otis in the present author's group [62,63].

*Figure 6: Photos from the first CALPHAD meeting in 1973 in Kaufman's home, (a) Hillert (1st from right), Cahn (2nd from right), Spencer (3rd from right), and Kaufman (showing back of his head); (b) Hillert (left) playing table tennis; (c) Kaufman (facing the camera) and his wife (1st from right).*

## 7 First and second derivatives of internal energy and free energy

### 7.1 First derivatives and definition of chemical potential

In the present author's textbook [37], he followed Hillert's procedure, but introduced the variable compositions in Eq. 3 and Eq. 4

$$dU = dQ - PdV + \sum_{i=1}^{c} H_i dN_i \qquad \textit{Eq. 8}$$

$$dS = \frac{dQ}{T} + \sum_{i=1}^{c} S_i dN_i + d_{ip}S \qquad \textit{Eq. 9}$$

$$S_i = \left(\frac{\partial S}{\partial N_i}\right)_{dQ=0,d_{ip}S=0,N_{j\neq i}} \qquad \textit{Eq. 10}$$

where $H_i$ and $S_i$ are the partial derivatives of total enthalpy and total entropy with respect to the moles of component $i$ under the adiabatic and no-work condition for the former and the adiabatic and equilibrium conditions for the latter, respectively.

However, soon after the book was published, the present author realized a problem with Eq. 3 and Eq. 8. Under the adiabatic, constant volume, and equilibrium conditions, they become

$$dU = H_m dN \qquad \textit{Eq. 11}$$

$$dU = \sum_{i=1}^{c} H_i dN_i \qquad \textit{Eq. 12}$$

The two sides of the equations are inconsistent. In writing the overview paper based on his Hume-Rothery lecture [64], the present author replaced $H_i$ by $U_i$ and obtained the combined law as follows

$$dU = dQ - PdV + \sum_{i=1}^{c} U_i dN_i = TdS - PdV + \sum_{i=1}^{c} \mu_i dN_i - Td_{ip}S \qquad \textit{Eq. 13}$$

$$\mu_i = U_i - TS_i \qquad \textit{Eq. 14}$$

where Eq. 14 gives the definition of chemical potential for the *first time* in the literature [64]. It is evident that the chemical potential of a component thus defined is related to the partial entropy of the component defined by Eq. 10. The inconsistency shown by Eq. 11 and Eq. 12 are thus resolved.

It is important to note that both $U_i$ and $\mu_i$ are the first derivative of internal energy with respect to the moles of the component, but under different conditions as shown below

$$U_i = \left(\frac{\partial U}{\partial N_i}\right)_{V,dQ=0,N_{j\neq i}} \qquad \textit{Eq. 15}$$

$$\mu_i = \left(\frac{\partial U}{\partial N_i}\right)_{S,V,N_{j\neq i},d_{ip}S=0} \qquad \textit{Eq. 16}$$

The conditions in Eq. 16 require a heat exchange between the system and its surroundings as derived from Eq. 9 as follows

$$dQ = -TS_i dN_i \qquad \textit{Eq. 17}$$

This means that the system must release this amount of heat reversibly to the surroundings when evaluating the chemical potential of the component. Otherwise, the derivative would be the partial internal energy of the component as shown by Eq. 15, not the chemical potential by Eq. 16.

The combined law and chemical potential in terms of Hillert thermodynamics and Gibbs energy with only the work by hydrostatic pressure is written as follows

$$dG = -SdT + VdP + \sum_{i=1}^{c} \mu_i dN_i - Td_{ip}S \qquad \textit{Eq. 18}$$

$$\mu_i = \left(\frac{\partial G}{\partial N_i}\right)_{T,P,N_{j\neq i},d_{ip}S=0} \qquad \textit{Eq. 19}$$

As $\mu_i$ is defined by Eq. 14, Eq. 19 is shown below to give the same result using the definition $G = H - TS$ with its independent variables kept constant, i.e., $T, P, N_{j\neq i}, d_{ip}S = 0$, which are omitted for brevity

$$\frac{\partial G}{\partial N_i} = \frac{\partial(H - TS)}{\partial N_i} = \frac{\partial Q + U_i \partial N_i}{\partial N_i} - \frac{\partial Q + TS_i \partial N_i}{\partial N_i} = U_i - TS_i \qquad \textit{Eq. 20}$$

where the following equation and Eq. 10 are used under the same conditions

$$dH = TdS + VdP + \sum_{i=1}^{c} \mu_i dN_i - Td_{ip}S \qquad \textit{Eq. 21}$$

It should be emphasized that $S_i$ and $U_i$ are defined by Eq. 10 and Eq. 15, respectively. It is evident that Eq. 20 can be applied to any free energy defined through the Legendre transformation as shown in Section 8.3.

Furthermore, the entropy production of each independent internal process can be divided into four actions: (1) heat generation $(d_{ip}Q)$, (2) consumption of some components as reactants $(dN_{r,j})$ with entropy $S_j$, (3) formation of some components as products $(dN_{p,k})$ with entropy $S_k$, and (4) reorganization of its configurations $(d_{ip}S^{config})$, as follows [5,65],

$$d_{ip}S = \frac{d_{ip}Q}{T} - \sum_j S_j dN_{r,j} + \sum_k S_k dN_{p,k} + d_{ip}S^{config} = \frac{D}{T} d\xi \qquad \textit{Eq. 22}$$

The reorganization of configurations is related to the information change of the internal process or the system with contributions from all internal processes in the system [65].

## 7.2 Second derivatives and cross phenomena

Second derivatives of internal or free energy are the first derivatives among potentials and molar quantities. The stability of a system is described by the derivative between a potential and its conjugate molar quantity, i.e.,

$$\frac{\partial Y_a}{\partial X_a} > 0 \qquad \textit{Eq. 23}$$

where the subscripts are used to denote that they are internal variables and are defined by the external variables when the system is in equilibrium with its surroundings. The limit of stability is reached with Eq. 23 equal to zero, and its inverse diverges positively

$$\frac{\partial X_a}{\partial Y_a} = +\infty \qquad \textit{Eq. 24}$$

However, there are no such requirements for derivatives between non-conjugated variables, and the divergence can thus be either positive or negative

$$\frac{\partial X_a}{\partial Y_b} = \frac{\partial X_b}{\partial Y_a} = \pm\infty \qquad \textit{Eq. 25}$$

Such an example is thermal expansion as follows

$$\frac{\partial V}{\partial T} = \frac{\partial S}{\partial(-P)} = \pm\infty \qquad \textit{Eq. 26}$$

Consequently, there are many examples where thermal expansion becomes negative under certain conditions such as water below 4 °C and invar alloys in certain temperature-pressure ranges and can be predicted by the zentropy theory discussed in Section 7.3 [4,66].

Ågren [67] used extensively the derivatives between chemical potential and compositions to extend the single diffusion mobility of a component in the lattice-fixed frame of reference to a vector of intrinsic diffusivity in the lattice-fixed frame of reference and a vector of the chemical or interdiffusion diffusivity in the volume-fixed frame of reference [68]. Both the intrinsic and chemical diffusivity coefficients are related to the mobility and are not independent kinetic coefficients [6]. Höglund and Ågren [69] extended the simulations to thermodiffusion of carbon, i.e., carbon diffusion driven by temperature gradient through the commonly used heat of transport, which could be connected to the derivative of chemical potential of carbon to temperature as discussed by the present author [5].

The present author started to look into thermodiffusion during his sabbatical leave with Murch and his team through molecular dynamics (MD) simulations [70–74]. It soon became clear that some fundamental information was missing. As discussed above, each diffusion component has only one independent diffusivity in the lattice-fixed frame of reference, and the experimentally measured chemical diffusivities are due to the derivatives of chemical potential of the diffusion component with respect to other components in the system [67,68]. By the same token, the experimentally measured or computationally simulated diffusion flux due to temperature gradient should be related to the derivative of chemical potential to temperature, including thermoelectricity that should be related to the derivative of the chemical potential of electrons to temperature. With the chemical potential of electrons evaluated from the Fermi-Dirac distribution as a function of temperature, the present author's team accurately predicted the Seebeck coefficients of several n- and p-type thermoelectric materials without any assumptions of the electron-scattering mechanism and with no fitting parameters [75,76].

The present author investigated further the phenomenological flux equations proposed by Onsager [43,77] and found some fundamental issues related to Onsager's reciprocal relations derived from the microscopic reversibility assumption in addition to the flux equations themselves [5]. Based on Hillert thermodynamics, the present author showed that the flux of a molar quantity is proportional *only* to the gradient of its conjugate potentials as there are no cross terms in the combined law, and all cross phenomena are due to the dependence of this potential on its conjugate molar quantities and other independent potentials and molar quantities [5–7]. The flux equation is thus written as follows

$$J_{\xi_j} = -L_{\xi_j} \nabla Y_{\xi_j} \qquad \textit{Eq. 27}$$

where $L_{\xi_j}$ is the kinetic coefficient for the change of the internal molar quantity $\xi_j$. The entropy production rate of the internal process can be written as

$$\frac{T d_{ip}\dot{S}_j}{V} = L_{\xi_j} \left(\nabla Y_{\xi_j}\right)^2 \qquad \textit{Eq. 28}$$

$\nabla Y_{\xi_j}$ is related to the gradients of its conjugate molar quantity and all other independent variables with some of them being molar quantities and some of them being potentials as follows

$$\nabla Y_{\xi_j} = \frac{\partial Y_{\xi_j}}{\partial \xi_j} \nabla \xi_j + \sum_{\xi_k \neq \xi_j} \frac{\partial Y_{\xi_j}}{\partial Y_{\xi_k}} \nabla Y_{\xi_k} + \sum_{\xi_l \neq \xi_j, \xi_k} \frac{\partial Y_{\xi_j}}{\partial \xi_l} \nabla \xi_l \qquad \textit{Eq. 29}$$

where the first and second summations represent independent potential and molar quantity variables, respectively. Eq. 27 can thus be further expanded as follows

$$J_{\xi_j} = -L_{\xi_j} \nabla Y_{\xi_j} = -L_{\xi_j} \left( \frac{\partial Y_{\xi_j}}{\partial \xi_j} \nabla \xi_j + \sum_{\xi_k \neq \xi_j} \frac{\partial Y_{\xi_j}}{\partial Y_{\xi_k}} \nabla Y_{\xi_k} + \sum_{\xi_l \neq \xi_j, \xi_k} \frac{\partial Y_{\xi_j}}{\partial \xi_l} \nabla \xi_l \right) \qquad \textit{Eq. 30}$$

It is noted that $L_{\xi_j}$ would also depend on all independent variables, i.e. $\xi_j$, $Y_{\xi_k}$, and $\xi_l$.

The derivatives in two summations in Eq. 30 represent the cross phenomena in the system [5] that Onsager intended to describe [43,77] and are listed in Table 1 and Table 2, respectively. Both tables are symmetric due to the Maxwell relations based on the second derivatives of internal or free energy. The diagonal quantities in Table 1 (marked in red color) are between conjugate variables and are positive for a stable system, while the off-diagonal quantities are between non-conjugate variables and can be either positive or negative as discussed above. Most of the quantities in Table 1 are well known physical properties except those in the last row and last column (in *italic text*), which are related to chemical reactions where the amount of a component changes with respect to potentials. The present author assigned some names to the quantities in the last row, but the names for those in the last column remain to be designated. The derivatives between potentials in Table 2 are less discussed in the literature though Gibbs [40,41] presented a number of them in connection with equilibria involving solid and interfaces between phases such as $\frac{\partial \mu_i}{\partial T}$, $\frac{\partial \mu_i}{\partial P}$, and $\frac{\partial P}{\partial T}$. These two tables demonstrate the usefulness of the Maxwell relation that Hillert was pondering in his lectures and include various cross phenomena such as thermoelectricity, thermodiffusion, electromigration, electrocaloric effect, and electromechanical effect [5–7].

*Table 1: Physical quantities related to first directives between molar quantities (first column) and potentials (first row)* [7] *shown by the derivatives in the second summation in Eq. 29. The quantities in red color are derivatives between conjugate variable and are positive in a stable system. The quantities in italic text are less discussed in the literature with new names designed by the present author.*

*Table 2: Cross phenomenon coefficients represented by first derivatives between potentials* [7] *as shown by the derivatives in the first summation in Eq. 29.*

### 7.3 Total entropy of a system by zentropy theory

Experimentally, entropy is obtained by the integration of measured heat capacity over temperature from 0 K to finite temperature using the 3rd law of thermodynamics, which is commonly referred as Clausius entropy or thermodynamic entropy. Theoretically, entropy is divided into two branches in terms of classical statistical mechanics and phonon vibrations. While the quantum statistical mechanics in principle captures both [78], it is intractable for practical applications. In statistical mechanics developed by Gibbs [79] in 1901 before quantum mechanics was developed, the entropy among configurations is related to the probability of each configuration. While in DFT-based quantum mechanics, the entropy due to thermal electrons and phonon vibrations is considered for a stable configuration and is often focused on the ground-state configuration. It is postulated that the total entropy of the system is composed of the entropies of individual configurations and the statistical entropy among all configuration and can be written as follows [65]

$$S = \sum_{k=1}^{m} p^k S^k - k_B \sum_{k=1}^{m} p^k ln p^k = \int_0^T \frac{C_P}{T} dT \qquad \textit{Eq. 31}$$

where $p^k$ and $S^k$ are the probability and entropy of configuration $k$, respectively, and $C_P$ is the heat capacity of the system. Eq. 31 represents the coarse graining of multiscale entropy [65] and was recently termed as zentropy theory [4]. It is noted that the first summation in Eq. 31 comes from the quantum mechanics and may thus be termed as quantum entropy, while the second summation is from Gibbs statistical mechanics and commonly called Gibbs entropy which is also called Shannon information entropy in the literature [80,81]. It is thus evident that the difference between the Clausius (thermodynamic) and Gibbs (Shannon) entropies is the quantum entropy, which seems related to the negentropy or "negative entropy" in the literature [82,83].

When each configuration is a pure quantum configuration without any additional internal degrees of freedom, the first summation in Eq. 31 would be zero, and the equation returns to Gibbs statistical mechanics. In the state-of-the-art DFT-based solutions to quantum mechanics, the configuration at zero K is the ground-state configuration in each system, and other metastable configurations are the non-ground-state symmetry-broken configurations in terms of the internal degrees of freedom of the ground-state configuration. Helmholtz energy of each configuration can then be predicted by DFT-based calculations [3]. Applying Eq. 31 to a system of canonical ensemble under constant NVT, the zetrnopy statistical mechanics of a system is obtained as follows [84]

$$F = \sum_{k=1}^{m} p^k E^k - TS = \sum_{k=1}^{m} p^k F^k + k_B T \sum_{k=1}^{m} p^k ln p^k \qquad Eq.\ 32$$

$$Z = e^{-\frac{F}{k_B T}} = \sum_{k=1}^{m} Z^k = \sum_{k=1}^{m} e^{-\frac{F^k}{k_B T}} \qquad Eq.\ 33$$

$$p^k = \frac{Z^k}{Z} = e^{-\frac{F^k - F}{k_B T}} \qquad Eq.\ 34$$

The only difference between Gibbs statistical mechanics and zentropy statistical mechanics is the replacement of the total energy of each configuration in the former by its free energy in the latter. The successful prediction of properties of magnetic materials including accurate predictions of magnetic phase transitions and other emergent properties without model parameters were reviewed by the present author recently [5–7,64,85]. It is being applied to transitions in ferroelectric materials [86] and superconductors [5,87].

## 8 Some challenging and confusing concepts in thermodynamics

### 8.1 Reference states

When the present author was at USTB, he asked undergraduate students what the most challenging concept was in thermodynamics, and the reference state was on the top of the list. While the present author was able to explain that the total energy in terms of $E = mc^2$ with $m$ being the mass and $c$ being the speed of light is too big when the chemical bonding energy between atoms is of interest so a more meaningful reference value of total energy or enthalpy is needed, but its arbitrariness hindered the full appreciation of the concept. On the other hand, the reference state for a chemical activity was purely for the convenience of experimental measurements. The connection between these two reference states and their usefulness were hard for students to comprehend.

It was at KTH that the present author realized the importance of reference state in building a large thermodynamic database with multiple components and phases. In the practice of the CALPHAD modeling of thermodynamics [88–90], the reference state is defined by the stable phase of a pure element at $T = 298.15\ K$ and $P = 101325\ Pa$ with its enthalpy equal to zero, referred as the stable element reference state (SER) [91]. This agreed SER state enabled various groups around the world to develop independently thermodynamics databases and combine them together to efficiently create larger databases. For chemical activities, the reference state can be changed to the experimental conditions so the chemical activities of pure elements equal to 1 for better graphic representations. The change of those reference states can be done easily in computer programs such as Thermo-Calc that the present author used for all calculations at KTH.

### 8.2 Vapor, partial, and total pressures

As mentioned in Section 2, these concepts played an important role in the admission of the present author to the graduate program in USTB. Even though their definitions are straightforward, the relationships among them depend on how the system is defined, whether the gas phase is considered, and whether the system is in equilibrium, and could sometimes create confusion in performing calculations.

The total pressure represents how a system is constrained by its surroundings. Its value is the sum of the partial pressure of each species in the gas phase in the system, including inert gas species. While the vapor pressure represents the pressure of a species in the gas phase when the species is equilibrated between the gas and liquid or solid phases, so its chemical potential or chemical activity is the same in all phases. If the vapor pressure of a species is larger than its partial pressure, the species evaporates into the gas phase. If its vapor pressure is lower than its partial pressure, the species condenses into the liquid or solid phase. If the vapor pressure equals to the total pressure, the system boils or sublimates.

For equilibrium calculations in a closed system without the gas phase present, the pressure can be set to an arbitrary value to fix one independent variable of the system. When the gas phase is present, if the total pressure is fixed, the volume of the system becomes a dependent variable, and vice versa. Furthermore, the partial pressure of each species equals to its vapor pressure in the system, which can be used to simulate the maximum chemical transport rate with a gas flow through the system [37]. To avoid the change of the liquid or solid phase compositions due to the different evaporation rates of species, the calculation can be made by fixing the gas phase with zero amount and unfixing the pressure, and the resulting total pressure is thus the sum of the vapor pressures of all species. This was how the $T - P$ phase diagram of the Mg-B binary system [8] was calculated where the all the three pressures are nearly the same due to the gas phase being almost pure Mg.

For open systems such as the growth of thin films in a molecular-beam epitaxy (MBE) chamber [9,10,92,93], partial pressures of precursors are controlled in experiments. These precursors react on the surface of a heated substrate to form a desired phase in an adsorption-controlled growth regime with low residual resistivities by mitigating defects resulting from cation non-stoichiometry [93]. For example, at low oxygen partial pressure and low temperature, the vapor pressures of species are low, the typical Ellingham diagrams can be obtained without the gas phase present in thermodynamics calculations. Two examples are shown in the low oxygen partial pressure ($P_{O_2}$) region of the $P_{O_2} - T$ phase diagrams in Figure 7 for (a) $SrRuO_3$ and (b) $CaRuO_3$, respectively [93]. The calculations were performed with solid phases only and without the gas phase. The vapor and partial pressures of each component are the same and represented by the activity in the system. The Gibbs phase rule by Hillert [1,2] stipulates that three-phase equilibria are invariant for a closed system under constant $T$ and $P$, i.e.,

$$v = c + 2 - p - n_s = 3 - p \qquad \textit{Eq. 35}$$

with $p$ being the number of phases in equilibrium, $c = 3$ for the number of independent components, and $n_s = 2$ for the number of potentials fixed, i.e., $T$ and $P$.

For $SrRuO_3$ and $CaRuO_3$, at $P_{O_2}$ higher than the reaction $Ru(solid) \rightarrow RuO_x(gas)$, the gas phase becomes stable, and the partial pressure of Ru ($P_{Ru}$) is lower than its vapor pressure ($P_{Ru}^v$). Consequently, $P_{Ru}$ needs to be fixed in the calculations instead of $P$, and the system becomes an open system. The Gibbs phase rule is the same as Eq. 35 with $n_s = 2$ for fixed $T$ and $P_{Ru}$, and the sum of partial pressures of all speciates gives the total pressure. Since the gas phase is stable, the invariant three-phase equilibria are with two solid phases instead of three discussed above. The corresponding phase diagram is shown in the upper high $P_{O_2}$ regions in Figure 7 for (a) $SrRuO_3$ and (b) $CaRuO_3$, respectively, separated from the low $P_{O_2}$ by the reaction $Ru(solid) \rightarrow RuO_x(gas)$. We termed this combined phase diagram as thermodynamics of MBE (TOMBE) diagram.

*Figure 7: Thermodynamics of MBE (TOMBE) illustrating the adsorption-controlled growth window for (a) $Sr_{n+1}Ru_nO_{3n+1}$ and (b) $Ca_{n+1}Ru_nO_{3n+1}$ phases. The light green shaded regions in (a) and (b) are the adsorption-controlled growth windows for $SrRuO_3$ and $CaRuO_3$, respectively [93]. The cyan lines show the equivalent oxidation potential for ozone partial pressures ranging from $10^{-11}$ to $10^{-5}$ Torr (1 Torr=133Pa). An excess ruthenium flux of $1.95x10^{17}$ atoms/$m^2$s and $3.2x10^{17}$ atoms/$m^2$s was used in the thermodynamic calculations for (a) and (b), respectively. Reproduced under a Creative Commons Attribution (CC BY) license.*

## 8.3 Molar quantity and potential vs. extensive and intensive variables, and chemical

**potential**

As in all textbooks before him, Hillert [1,2] started the discussion of state variables in terms of extensive and intensive variables, but soon switched to molar quantities and potentials and used them for the rest of his book. This was due to the ambiguity of intensive variables which can be either potentials or normalized molar quantities. As shown in Eq. 2 and Eq. 7, all molar quantities and the internal energy are extensive variables plus all free energies derived from the internal energy, and they can all be normalized. While potentials are defined by the derivatives of the internal energy with respect to the molar quantities and cannot be further normalized. Furthermore, each potential has the same value everywhere in a heterogenous equilibrium systems where the heterogeneity is due to the heterogeneity of normalized molar quantities. Therefore, one should use potentials and molar quantities and abandon the use of extensive and intensive variables in thermodynamics as Hillert did.

Furthermore, it is critical to realize that partial entropy and partial internal energy are fundamentally different from normalized entropy and normalized internal energy. The partial quantity represents the partial derivative of a quantity, i.e., the slope or change rate of the quantity with respect to the moles of a component, while the normalized quantity is a simple arithmetic division of a quantity even though both the partial quantity and the normalized quantity have the same unit. They are thus both fundamentally and mathematically different from each other. Therefore, the normalized energy and free energies should not be called potentials as commonly done in almost all textbooks.

Molar Gibbs energy of a system, $G_m$ shown in Eq. 6, is worth of further discussion. It is tempting to consider $G_m$ as a potential, i.e., similar to an $Y^a$ in Eq. 2 or Eq. 7 as shown in many textbooks, due to its equivalent position as $T$ and $-P$ in Eq. 6. However, since the change of the size of a system in terms of $dN$ does not usually induce any internal processes and thus does not alter the state of the system, the definition of $G_m$ as a potential is thus without any practical usefulness. Furthermore, the chemical potential of each component is introduced in Eq. 2 and Eq. 7, which plays an essential role in defining the equilibrium state of the system as discussed in Section 7.1. It is thus both confusion and unnecessary to term $G_m$ as a potential as it does not play the role of potential that must have the same value everywhere in an equilibrium system with compositional heterogeneity, while $G_m$ does not have the same value everywhere in such a system. It is noted that for systems with one independent component, $G_m = \mu_A$, the concept of chemical potential is not useful due to no variation of compositions in the system, and $G_m$ or $\mu_A$ is usually treated as a dependent variable with $T$ and $-P$ as the two independent potentials as shown by Hillert [1,2]. Additionally, by the definition of $G_m = U_m - TS_m - (-P)V_m$ with $U_m$, $S_m$, and $V_m$ being molar internal energy, molar entropy and molar volume of the system, $G_m$ is an energy that represents portion of the internal energy of the system with the contributions due to entropy and volume removed, noting $-P$ defined as the conjugate potential of volume as shown by Hillert [1,2]. The usefulness of $G_m$ is because its natural variables, i.e., $T$, $-P$, and $x_i$, are variables that are usually controlled in experiments, and $G_m$ is minimized in systems with those natural variables being kept constant.

Now let us examine the chemical potential in more detail, which was added into Eq. 2 by Gibbs or Eq. 7 by Hillert without any discussions, resulting in significant difficulty in comprehending its physical meaning. The present author defines the chemical potential by Eq. 14 through partial entropy and partial internal energy defined by Eq. 10 and Eq. 15, respectively [64]. The significance of Eq. 14 is that the chemical potential of a component equals to its partial internal energy minus the contribution due to its partial entropy, i.e., representing only portion of the partial internal energy of the component in the system. One may attempt to equal it to the partial Helmholtz energy due to $F = U - TS$; however, this is not correct as the derivatives of Eq. 10 and Eq. 15 are calculated under different conditions than the partial Helmholtz energy, which is as follows

$$F_i = \left(\frac{\partial F}{\partial N_i}\right)_{T,V,N_{j\neq i},d_{ip}S=0} = \left(\frac{\partial U}{\partial N_i}\right)_{T,V,N_{j\neq i},d_{ip}S=0} - T\left(\frac{\partial S}{\partial N_i}\right)_{T,V,N_{j\neq i},d_{ip}S=0} \quad \text{Eq. 36}$$

It is evident that the conditions are different for the partial derivatives in Eq. 36 in comparison with those in Eq. 10 and Eq. 15.

Consequently, the chemical potential of a component is the derivative of the internal energy with respect to the moles of the component and is fundamentally and mathematically different from the molar Gibbs energy even though they have the same unit. Furthermore, the chemical potential of a component can be represented by any characteristic state functions or free energies if it has the moles of this component as one of its natural variables. It is shown by Eq. 20 that the chemical potential evaluated from Gibbs energy is the same as that from the internal energy. The same procedure can be performed for enthalpy and Helmholtz energy as follows with the common subscript of $N_{j\neq i}, d_{ip}S = 0$ omitted for clarity

$$\left(\frac{\partial H}{\partial N_i}\right)_{S,P} = \frac{\partial(U+PV)}{\partial N_i} = \frac{\partial Q - P\partial V + U_i\partial N_i + P\partial V}{\partial N_i} = \frac{(U_i - TS_i)\partial N_i}{\partial N_i} = U_i - TS_i \qquad \textit{Eq. 37}$$

$$\left(\frac{\partial F}{\partial N_i}\right)_{T,V} = \frac{\partial(U-TS)}{\partial N_i} = \frac{\partial Q - P\partial V + U_i\partial N_i - T\partial S}{\partial N_i} = \frac{(U_i - TS_i)\partial N_i}{\partial N_i} = U_i - TS_i \qquad \textit{Eq. 38}$$

This may seem trivial as one can directly obtain the following relations from the combined law with the same subscript of $N_{j\neq i}, d_{ip}S = 0$ omitted for brevity

$$\mu_i = \left(\frac{\partial U}{\partial N_i}\right)_{S,V} = \left(\frac{\partial H}{\partial N_i}\right)_{S,P} = \left(\frac{\partial F}{\partial N_i}\right)_{T,V} = \left(\frac{\partial G}{\partial N_i}\right)_{T,P} = U_i - TS_i \qquad \textit{Eq. 39}$$

However, the fact that they all equal to $U_i - TS_i$ with $U_i$ and $S_i$ from Eq. 10 and Eq. 15, respectively, has not been proved before. This outcome is of course not surprising due to the internal consistence of characteristic state functions. It is noted that as discussed in Section 4, $\mu_i$ in terms of the molar free energy is complicated and is given in Hillert's textbook [1,2].

### 8.4 Derivation of Gibbs-Duhem equation

In the papers by Gibbs [40,41] and most thermodynamic text books if not all including the ones by Hillert [1,2] and by the present author and Wang [37], the derivation of the Gibbs-Duhem equation starts as follows by expanding a small portion in a homogeneous equilibrium system

$$dU = TdS - PdV + \sum_{i=1}^{c} \mu_i dN_i = \left(TS_m - PV_m + \sum_{i=1}^{c} \mu_i x_i\right) dN \qquad \textit{Eq. 40}$$

$$U = \left(TS_m - PV_m + \sum_{i=1}^{c} \mu_i x_i\right) N = TS - PV + \sum_{i=1}^{c} \mu_i N_i \qquad \textit{Eq. 41}$$

The validity of this procedure has bothered the present author for decades. Recently, the present author realized the flaw in Eq. 40. The three terms in the first part of Eq. 40 represent three independent exchanges between the system and its surroundings, while the three terms in the second part of Eq. 40 combine them into one exchange, indicating that they are not independent. Furthermore, $dS$ contains contributions from $dN_i$ as shown by Eq. 4 and Eq. 9, and they can thus not be independent. Therefore, Eq. 40 is self-contradictory. More rigorous procedure is thus needed.

Let us perform a virtual experiment as Hillert often said in his class and consider that only exchange from the surrounding to the system is the addition of mass, which results in the following equation from the first part of Eq. 13,

$$dU = \sum_{i=1}^{c} U_i dN_i \qquad \text{Eq. 42}$$

To maintain constant $V$ according to the virtual experiment, the pressure of the system is increased as each component has a partial volume of $V_i$. To maintain the equilibrium between the system and its surroundings by equalizing the pressure, the volume of the system must increase, resulting in the loss of the internal energy of the system by the following amount

$$-PdV = -P\sum_{i=1}^{c} V_i dN_i = -Td_{ip}S \qquad \text{Eq. 43}$$

Inserting Eq. 43 into Eq. 42 and using Eq. 14 for the definition of $\mu_i$, one obtains

$$dU = \sum_{i=1}^{c} U_i dN_i - P\sum_{i=1}^{c} V_i dN_i = T\sum_{i=1}^{c} S_i dN_i + \sum_{i=1}^{c} \mu_i dN_i - P\sum_{i=1}^{c} V_i dN_i \qquad \text{Eq. 44}$$

When each component is added proportionally as specified by the virtual experiment, one obtains the second part of Eq. 40 as follows

$$dU = \left(T\sum_{i=1}^{c} S_i x_i + \sum_{i=1}^{c} \mu_i x_i - P\sum_{i=1}^{c} V_i x_i\right) dN = \left(TS_m - PV_m + \sum_{i=1}^{c} \mu_i x_i\right) dN \qquad \text{Eq. 45}$$

Eq. 41 is thus rigorously proven and can be used to derive the Gibbs-Duhem equation as follows

$$0 = -SdT - Vd(-P) - \sum_{i=1}^{c} N_i d\mu_i \qquad \textit{Eq. 46}$$

## 8.5 Gibbs phase rule and phase diagrams

Gibbs phase rule is presented by the first part of Eq. 35, i.e.,

$$\upsilon = c + 2 - p - n_s \qquad \textit{Eq. 47}$$

Hillert emphasized that the Gibbs phase rule concerns the number of potentials that can vary independently without changing the number of phases in equilibrium because the Gibbs-Duhem equation (Eq. 46) used to derive the Gibbs phase rule is about the relation between potentials only. However, in most textbooks, $\upsilon$ is loosely called degrees of freedom, resulting in considerable confusion in the literature [34] because the number of independent variables in a system is equal to c+2 and does not change with the number of phases in equilibrium. What changes is the number of potentials among the c+2 independent variables. From the Gibbs-Duhem relation, the maximum number of potentials among the c+2 independent variables is c+1 for a single-phase region. When there are c+2 phases in equilibrium, all potentials are fixed, and all c+2 independent variables must all be molar quantities. The changes of molar quantities only affect the relative amount of each phase and not the number of phases in equilibrium.

Hillert discussed potential, molar, and mixed phase diagrams with their axes being all potentials, all molar quantities, or mixture of potentials and molar quantities, respectively [2,94]. The Gibbs phase rule is strictly appliable to potential phase diagrams only though it could be modified for molar or mixed phase diagrams [2,94]. The number of axes of a potential phase diagram equals to $c + 2 - 1 = c + 1$ where 2 represents T and P. For systems under stress, electric, and magnetic fields, the number of potentials increases accordingly [95]. The c+1 geometric features in the potential diagram represent single phase regions, and the dimension of the geometric feature reduces by one with one additional phase added into the equilibrium until the dimension becomes zero for c+2 phases in equilibrium. It is noted that all geometric features are phase regions in a potential phase diagram, and there are no phase boundaries.

When one potential is changed to its conjugate molar quantity, the dimensions of phase regions with two and more phases expand by one, resulting in a mixed phase diagram with one axis being molar quantity. Phase boundaries are introduced between the single- and two-phase regions along with tie-lines that connect the two phases in equilibrium along the molar axis with two phases having different values of the molar quantity. The dimensions of phase regions lower than the dimensions of the phase diagram continues to increase with more potentials replaced by their conjugate variables until all axes of the phase diagram are molar quantities, and all phase regions have the same dimensions as the phase diagram, thus a molar phase diagram.

Due to the introduction of phase boundaries, the Gibbs phase rule is no longer applicable to mixed or molar phase diagrams, while the lever rule and the phase boundary rule can be applied as discussed in detail by Hillert [2,94]. During 1997 to 1998, the present author and Charlie Kuehmann at QuesTek Innovations LLC traveled across the US to give short courses on Thermo-Calc and Dictra with laptops packed in cases. Several professors commented that this is how phase diagrams should be taught in classes, which inspired the present author to apply for fundings from National Science Foundation for education [24] after he joined PSU and continue the short courses until 2011 [96].

## 8.6 Second law of thermodynamics

The fundamental challenge in discussing the validity of the second law of thermodynamics is the difficulty in quantitative evaluation of entropy and entropy production due to internal processes. Hillert [1,2] stated that "the second law is derived only for the whole system", which is valid for a system with one independent internal process which may consist of several dependent internal processes. For a system with multiple independent processes, every of them needs to obey the second law of thermodynamics with positive entropy production. The violation of second law of thermodynamics has been discussed in the literature since the second law of thermodynamics was formulated with the latest argument in the microscopic scale in the framework of fluctuation theorem [97,98]. It is probably due to the incomplete evaluation of all entropy contributions to the internal process [5–7].

One rigorous experiment was performed by Maillet et al. [99] who built a single-electron transistor two-level system with their free energies being equal and a single thermodynamic trajectory level coupled to a single heat bath. They demonstrated that the amount of work up to large fractions of $k_B T$ could be extracted with two carefully designed out-of-equilibrium driving cycles featuring kicks of the control parameter despite the zero free energy difference over the cycle. This seemed an apparent violation of the second law of thermodynamics [100–103]. However, Maillet et al. [99] pointed out that the requirement of an external intervention makes the second law of thermodynamics remain valid.

It seems that in most discussions related to the “violation of second law of thermodynamics”, the second summation in Eq. 31, i.e., the Gibbs (Shannon information) entropy, is not included, resulting in incomplete accounting of entropy production represented by $d_{ip}S^{config}$ in Eq. 22 so Eq. 22 becomes negative, thus a false conclusion on the violation of the second law of thermodynamics.

## 9 Summary

It has been a privilege for the present author to learn from Hillert for 35 years starting in 1987. Standing on Hillert’s shoulder, the present author has gained much deeper fundamental understanding of thermodynamics, broader appreciation of its complexity, and wider perspectives of its applications, which would not have been possible otherwise. The present author expresses his most sincere gratitude for such an extraordinary opportunity, for without it, the present author might still be stuck in the basin of equilibrium thermodynamics or improperly applying equilibrium thermodynamics to nonequilibrium systems.

The present paper further shows that the predictive thermodynamics is becoming possible through the integration of bottom-up DFT-based calculations and top-down statistical mechanics in terms of the zentropy theory. There are many interesting and challenging topics to apply Hillert nonequilibrium thermodynamics because everything around us and in the space is not at equilibrium. Some particularly exciting topics to the present author include superconductivity [5,87], black holes [6], information [64,65], and many other large and small systems [7].

## 10 Acknowledgements

The present author thanks all his current and former students and numerous collaborators over the years at Penn State and around the world as reflected by the co-authors in the references cited in this paper. The present review article covers research outcomes supported by multiple funding agencies over half century as reflected in acknowledgements of the references cited. The most recent ones including the Endowed Dorothy Pate Enright Professorship at the Pennsylvania State University, U.S. Department of Energy (DOE) Grant No. DE-SC0023185, DE-AR0001435, DE-NE0008945, and DE-NE0009288, U.S. National Science Foundation (NSF) Grant No. NSF-2229690, NSF-2226976, and NSF-2050069, and Office of Naval Research (ONR) Grant No. N00014-21-1-2608 and N00014-23-2721.

Table 1: Physical quantities related to first directives between molar quantities (first column) and potentials (first row) [7] shown by the derivatives in the second summation in Eq. 29. The quantities in red color are derivatives between conjugate variable and are positive in a stable system. The quantities in italic text are less discussed in the literature with new names designed by the present author.

| | *T*, Temperature | $\boldsymbol{\sigma}$, Stress | $\boldsymbol{E}$, Electrical field | $\boldsymbol{\mathcal{H}}$, Magnetic field | $\mu_i$, Chemical potential |
|---|---|---|---|---|---|
| *S*, Entropy | Heat capacity | Piezocaloric effect | Electrocaloric effect | Magnetocaloric effect | $\frac{\partial S}{\partial \mu_k}$ |
| $\boldsymbol{\varepsilon}$, Strain | Thermal expansion | Elastic compliance | Converse piezoelectricity | Piezomagnetic moduli | $\frac{\partial \varepsilon_{ij}}{\partial \mu_k}$ |
| $\boldsymbol{\theta}$, Electric displacement | Pyroelectric coefficients | Piezoelectric moduli | Permittivity | Magnetoelectric coefficient | $\frac{\partial D_i}{\partial \mu_k}$ |
| $\boldsymbol{B}$, Magnetic induction | Pyromagnetic coefficient | Piezomagnetic moduli | Electromagnetic coefficient | Permeability | $\frac{\partial B_i}{\partial \mu_k}$ |
| $N_j$, Moles | *Thermoreactivity* | *Stressoreactivity* | *Electroreactivity* | *Magnetoreactivity* | $\frac{\partial N_i}{\partial \mu_i}$, $\frac{\partial N_i}{\partial \mu_k}$, *Thermodynamic factor* |

*Table 2: Cross phenomenon coefficients represented by first derivatives between potentials* [7] *as shown by the derivatives in the first summation in Eq. 29.*

| | $T$, Temperature | $\boldsymbol{\sigma}$, Stress | $\boldsymbol{E}$, Electrical field | $\boldsymbol{\mathcal{H}}$, Magnetic field | $\mu_i$, Chemical potential |
|---|---|---|---|---|---|
| $T$ | 1 | $-\frac{\partial S}{\partial \boldsymbol{\varepsilon}}$ | $-\frac{\partial S}{\partial \boldsymbol{\theta}}$ | $-\frac{\partial S}{\partial \boldsymbol{B}}$ | $-\frac{\partial S}{\partial c_i}$ Partial entropy |
| $\boldsymbol{\sigma}$ | $\frac{\partial \boldsymbol{\sigma}}{\partial T}$, Thermostress | 1 | $-\frac{\partial \boldsymbol{\varepsilon}}{\partial \boldsymbol{\theta}}$ | $-\frac{\partial \boldsymbol{\varepsilon}}{\partial \boldsymbol{B}}$ | $-\frac{\partial \boldsymbol{\varepsilon}}{\partial c_i}$ Partial strain |
| $\boldsymbol{E}$ | $\frac{\partial \boldsymbol{E}}{\partial T}$, Thermoelectric | $\frac{\partial \boldsymbol{E}}{\partial \sigma}$, Piezoelectric | 1 | $-\frac{\partial \boldsymbol{\theta}}{\partial \boldsymbol{B}}$ | $-\frac{\partial \boldsymbol{\theta}}{\partial c_i}$ Partial electrical displacement |
| $\boldsymbol{\mathcal{H}}$ | $\frac{\partial \boldsymbol{\mathcal{H}}}{\partial T}$, Thermomagnetic | $\frac{\partial \boldsymbol{\mathcal{H}}}{\partial \boldsymbol{\sigma}}$, Piezomagnetic | $\frac{\partial \boldsymbol{\mathcal{H}}}{\partial \boldsymbol{E}}$, Electromagnetic | 1 | $-\frac{\partial B}{\partial c_i}$ Partial magnetic induction |
| $\mu_i$ | $\frac{\partial \mu_i}{\partial T}$ Thermodiffusion | $\frac{\partial \mu_i}{\partial \boldsymbol{\sigma}}$ Stressmigration | $\frac{\partial \mu_i}{\partial \boldsymbol{E}}$ Electromigration | $\frac{\partial \mu_i}{\partial \boldsymbol{\mathcal{H}}}$ Magnetomigration | $\frac{\partial \mu_i}{\partial \mu_j} = -\frac{\partial c_j}{\partial c_i} = \frac{\Phi_{ii}}{\Phi_{ji}}$ Crossdiffusion |

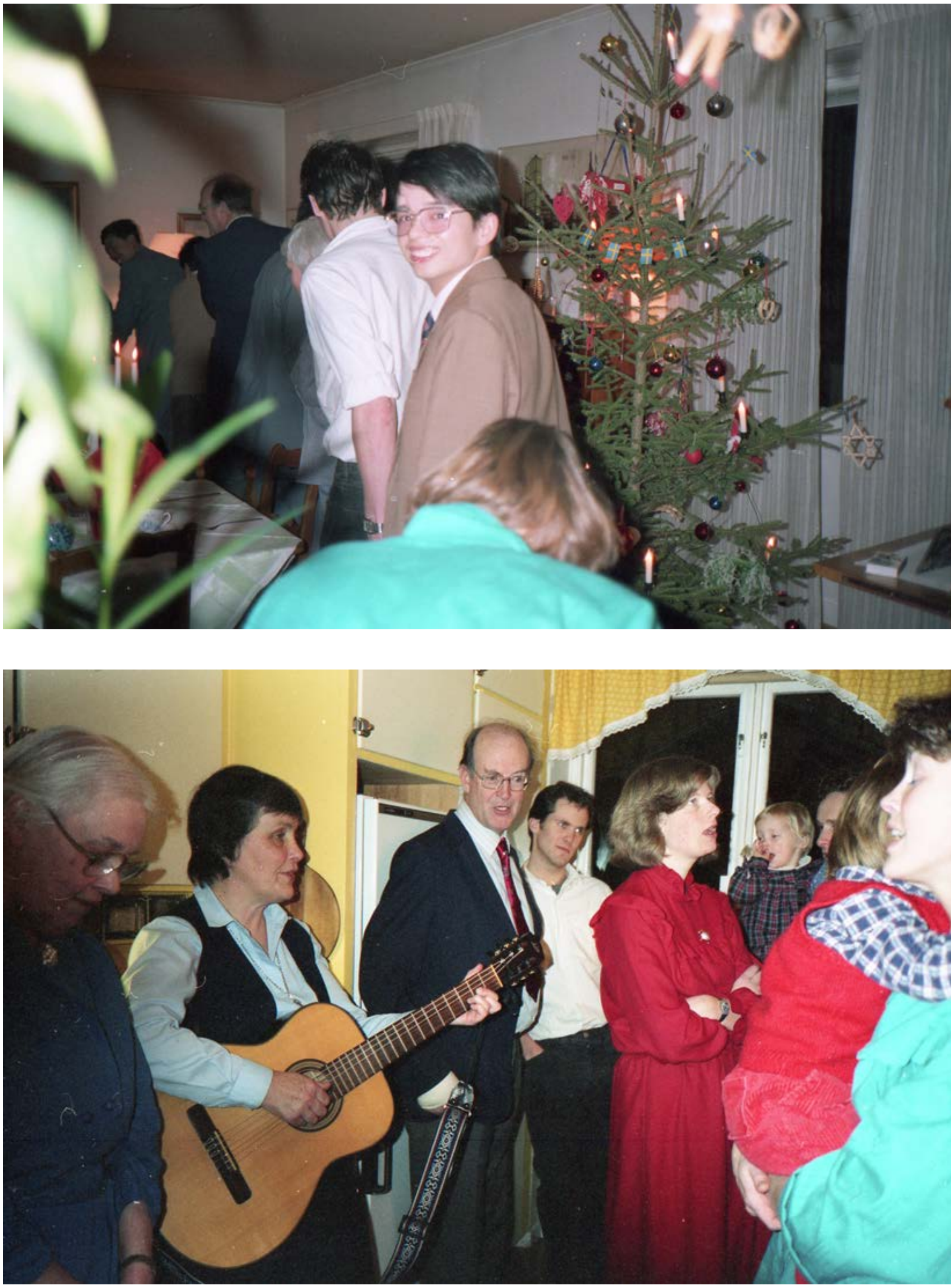

(a)

(b)

Figure 1: Christmas photos at Hillert's house in December 1987, (a) dancing around in the house, and (b) singing in the kitchen while someone was hiding an item for others to find.

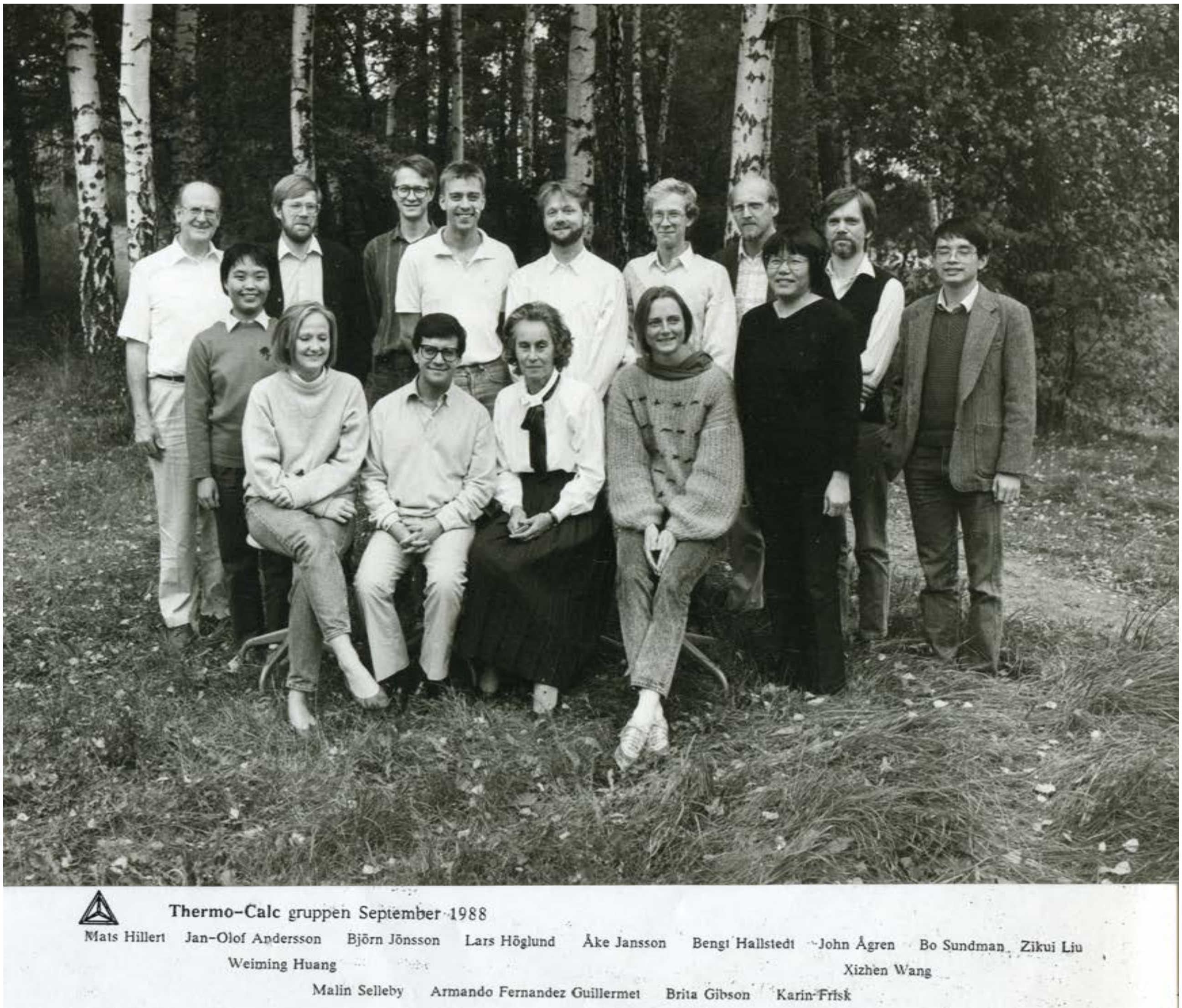
Thermo-Calc gruppen September 1988
Mats Hillert Jan-Olof Andersson Björn Jönsson Lars Höglund Åke Jansson Bengt Hallstedt John Ågren Bo Sundman Zikui Liu
Weiming Huang
Xizhen Wang
Malin Selleby Armando Fernandez Guillermet Brita Gibson Karin Frisk

(a)

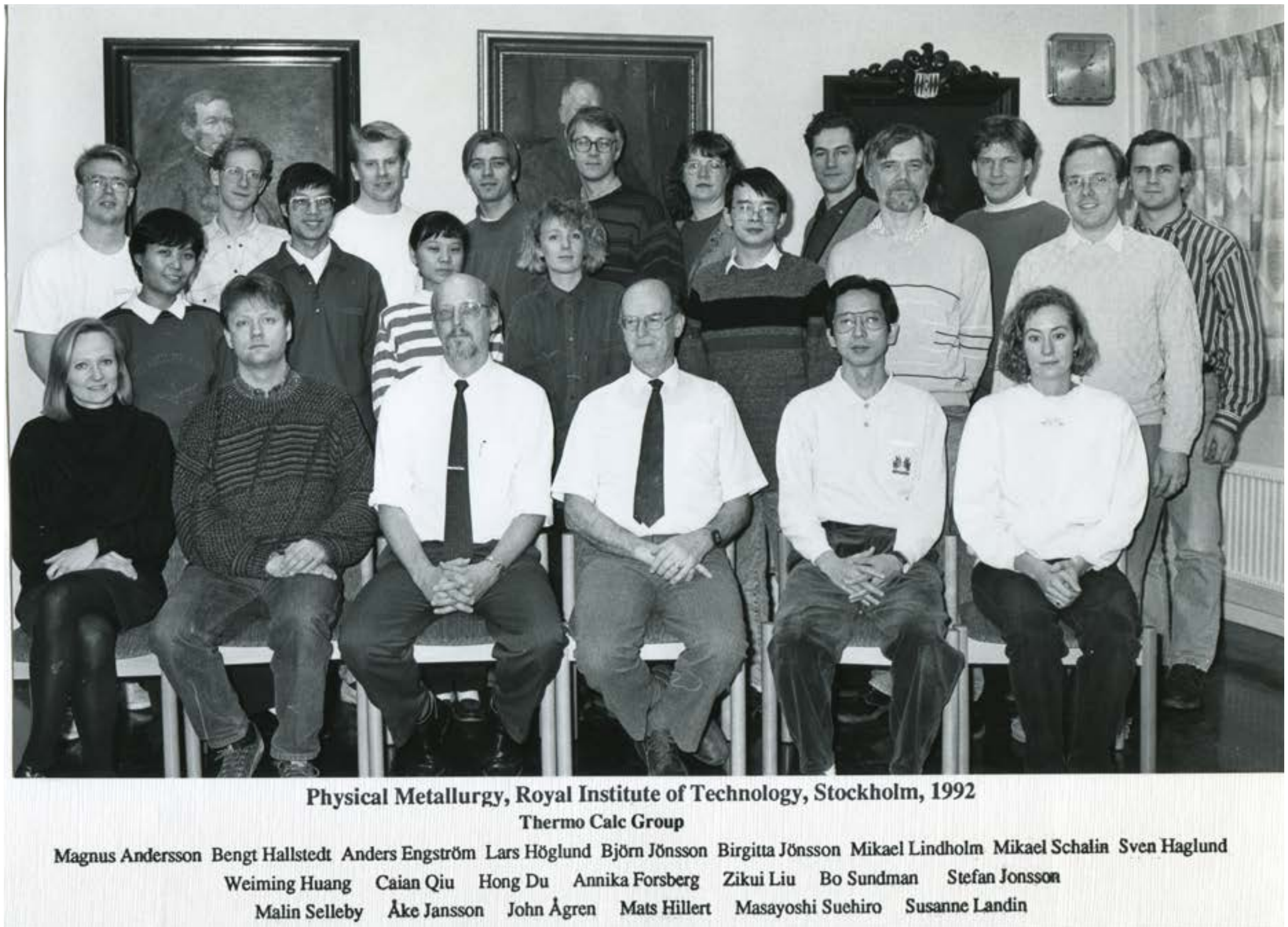

(b)

Figure 2: Group photos in (a) 1988 and (b) 1992, KTH, Sweden.

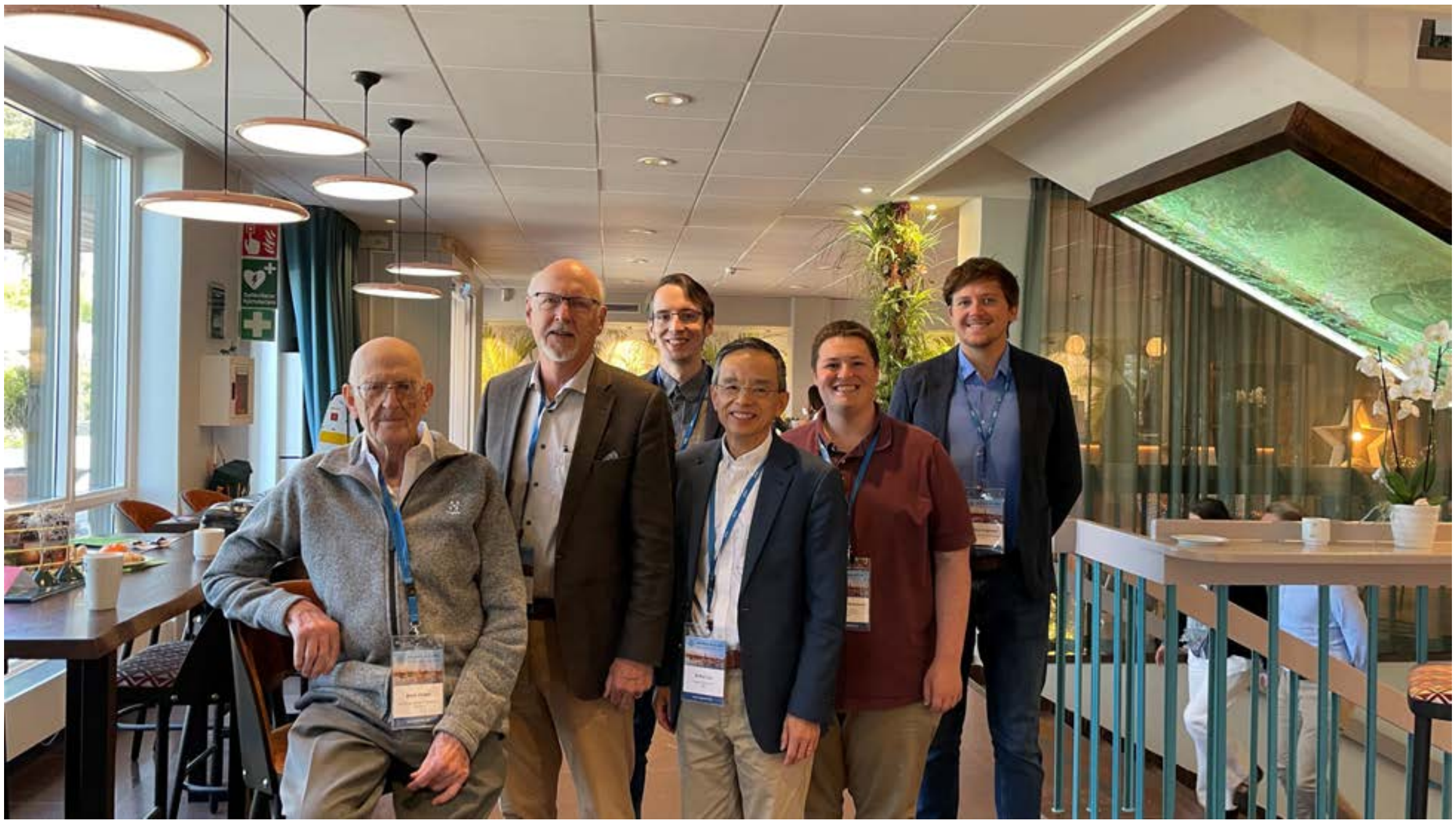

Figure 3: Group photo on May 23, 2022 at the CALPHAD conference in Lidingö, Sweden.

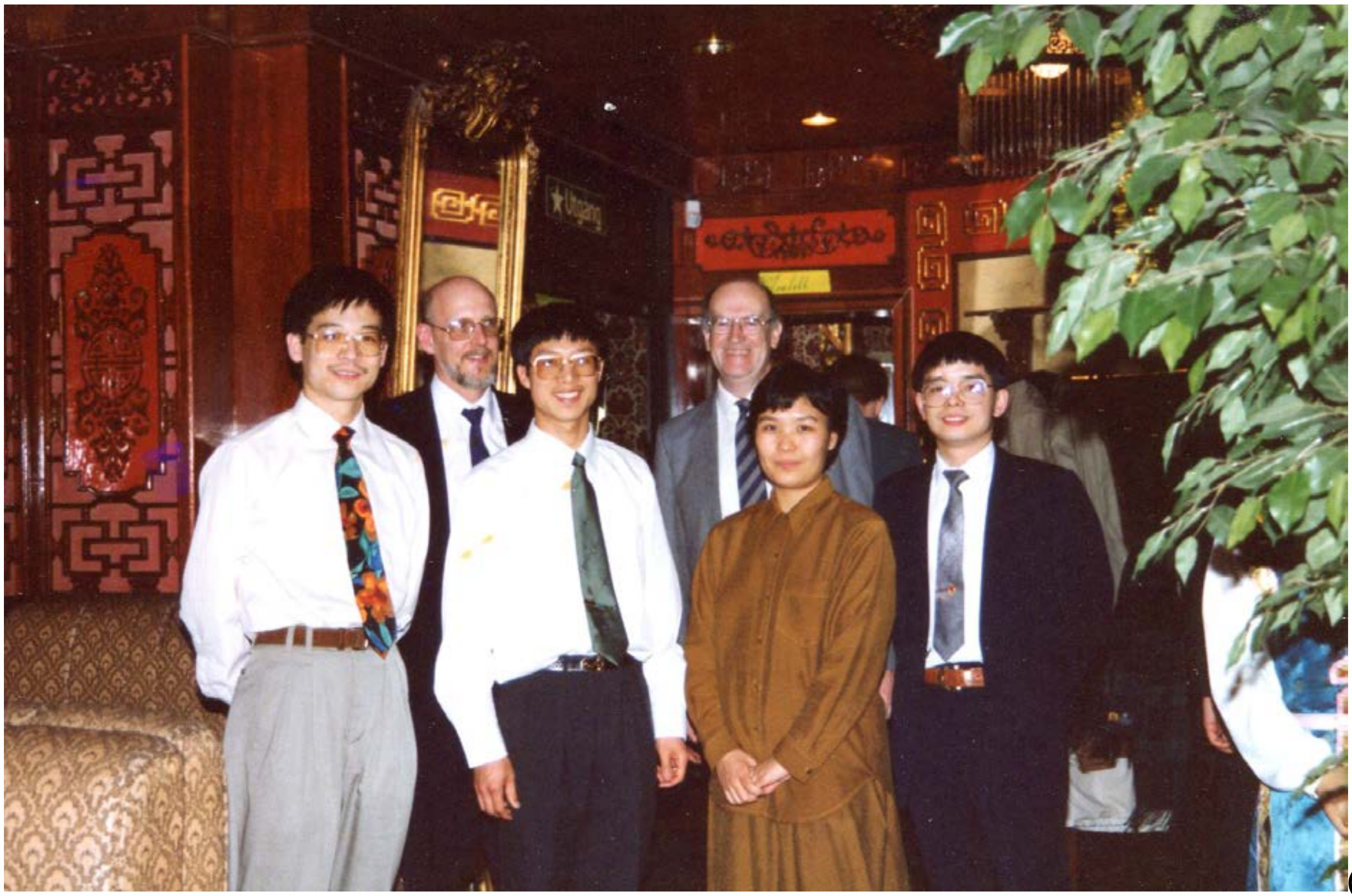

(a)

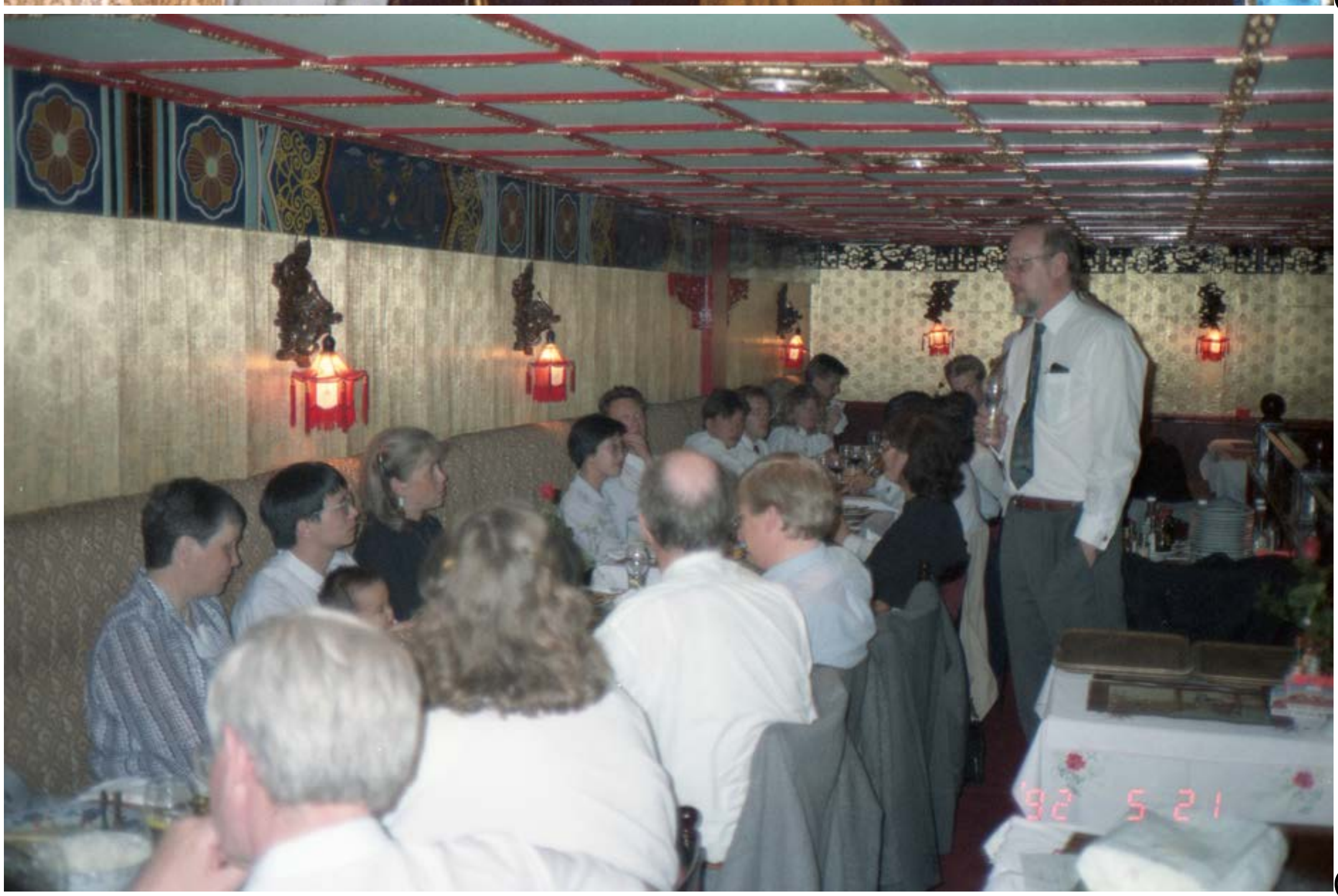

(b)

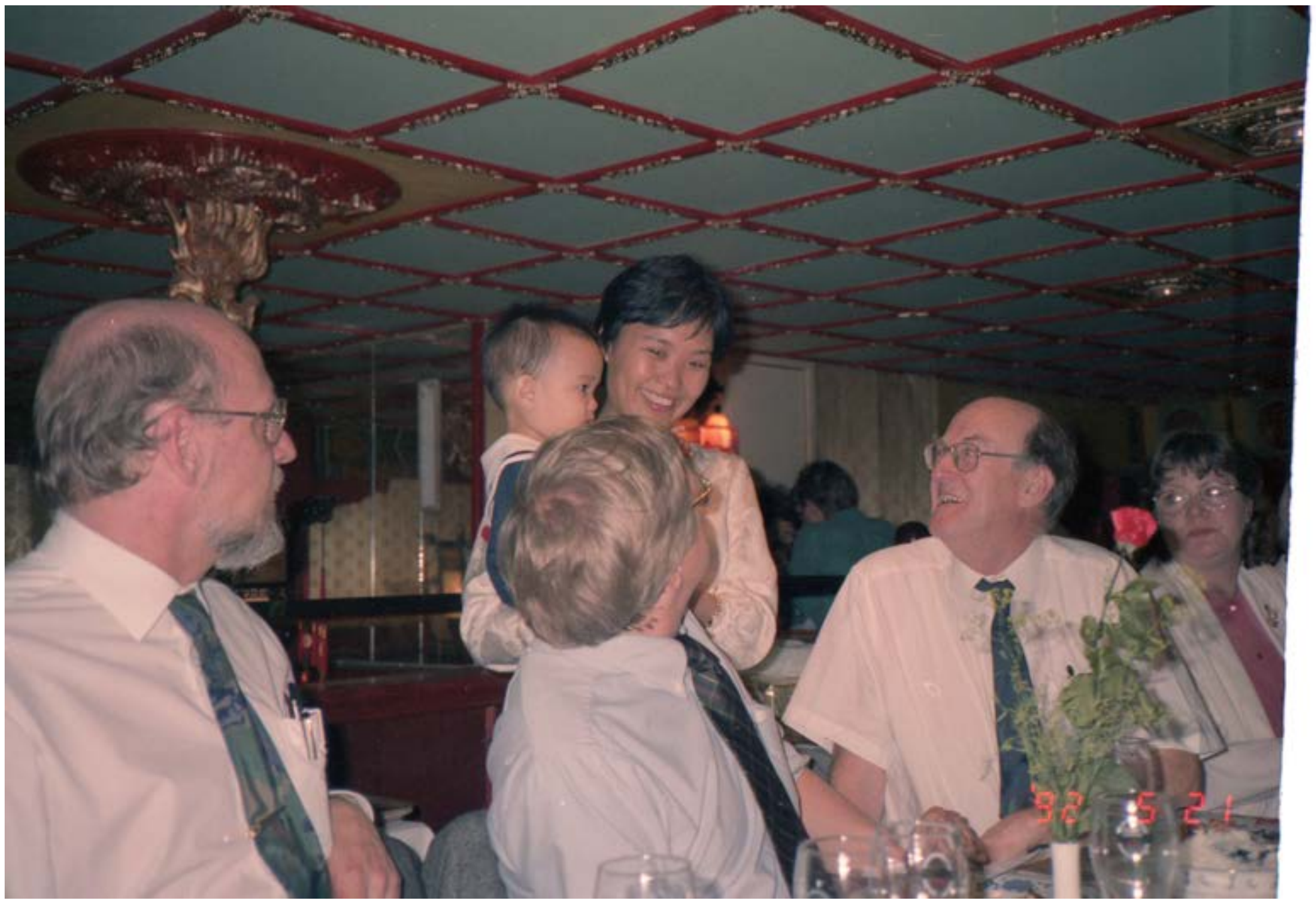
(c)

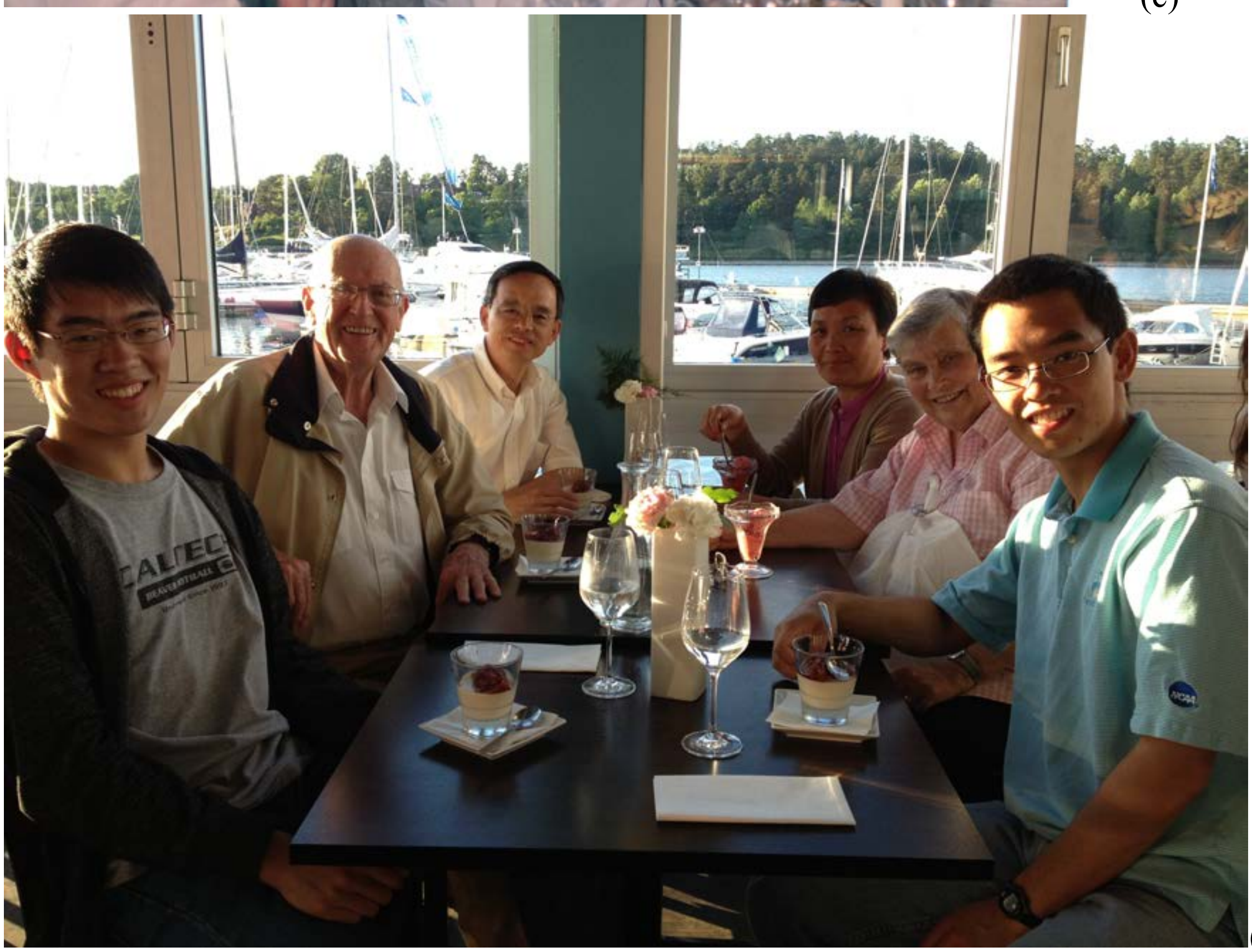
(d)

Figure 4: (a) Banquet after Weiming Huang's PhD defense on October 30, 1990, (b) & (c) Banquet after the present author's PhD defense on May 21, 1992, (d) Dinner on the island of Hillert's house during the visit to Sweden on June 17, 2013.

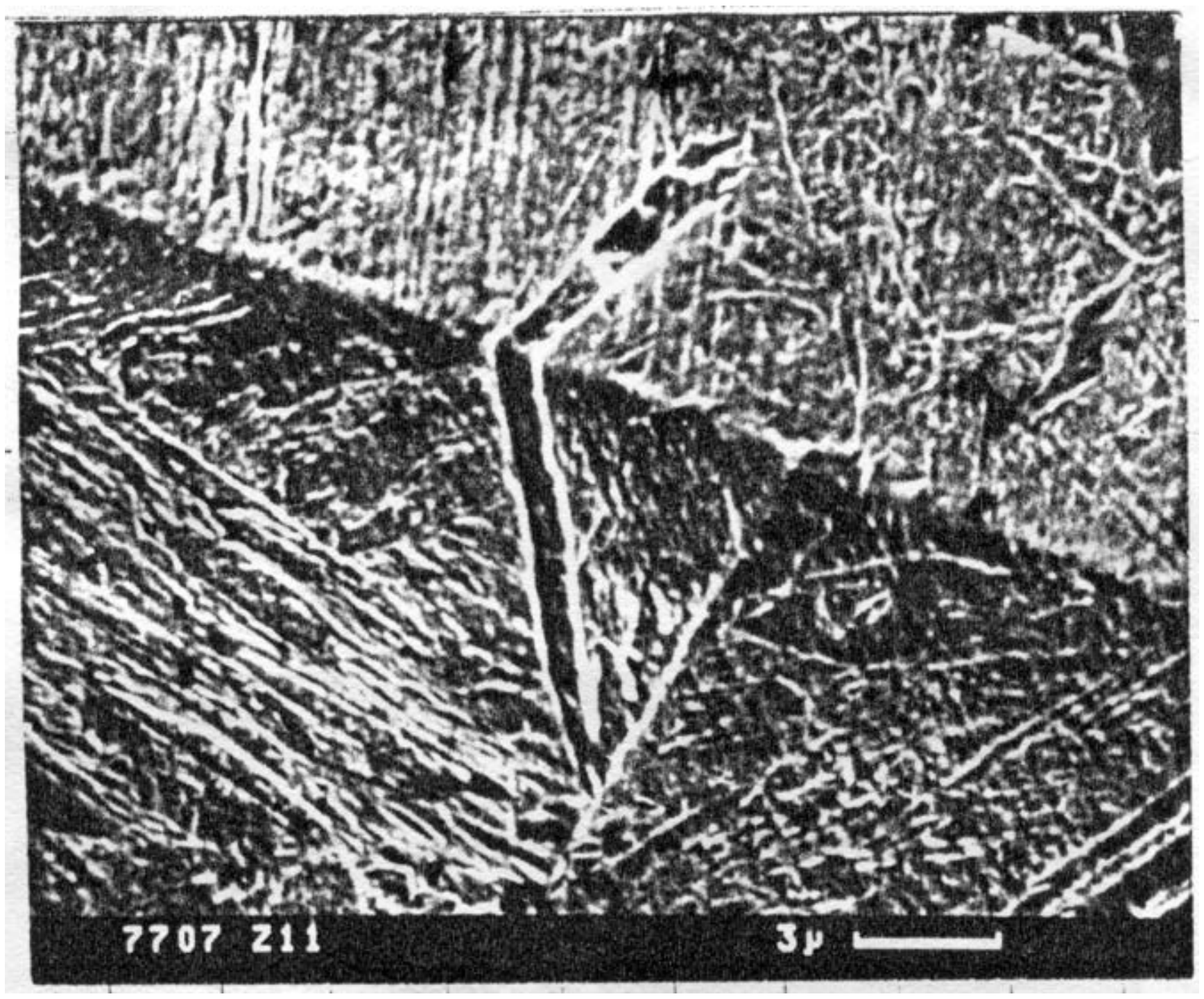

(a)

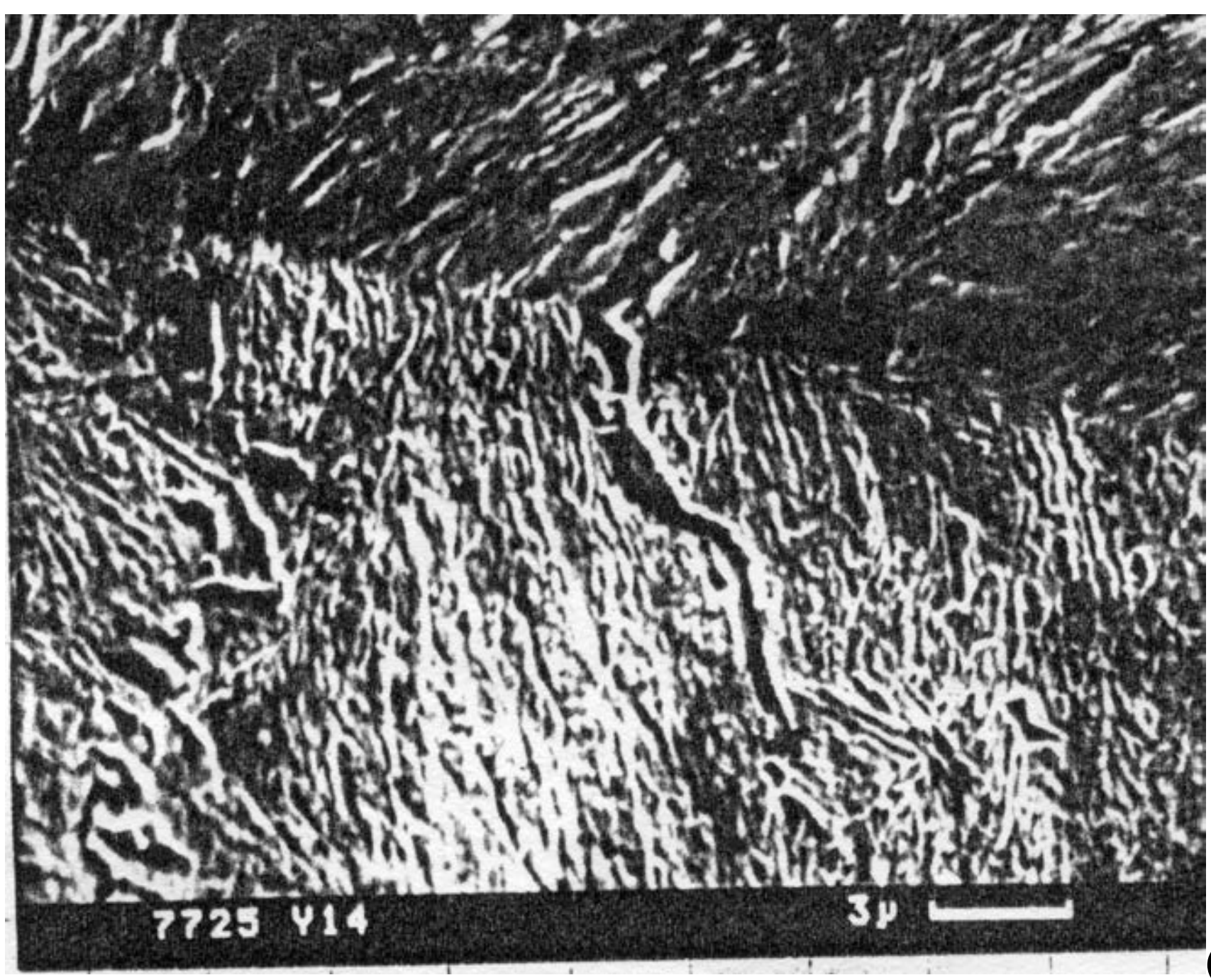

(b)

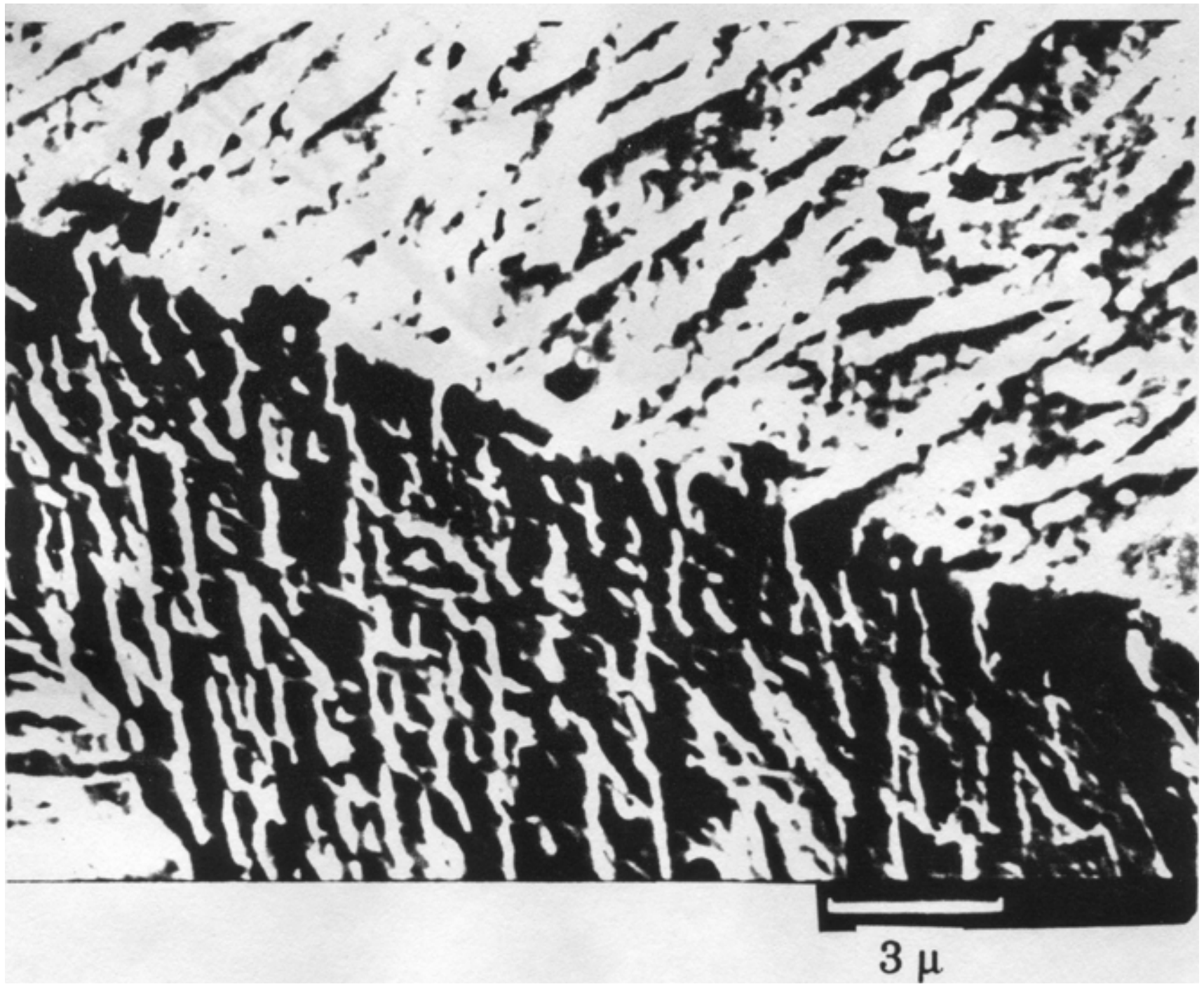

(c)

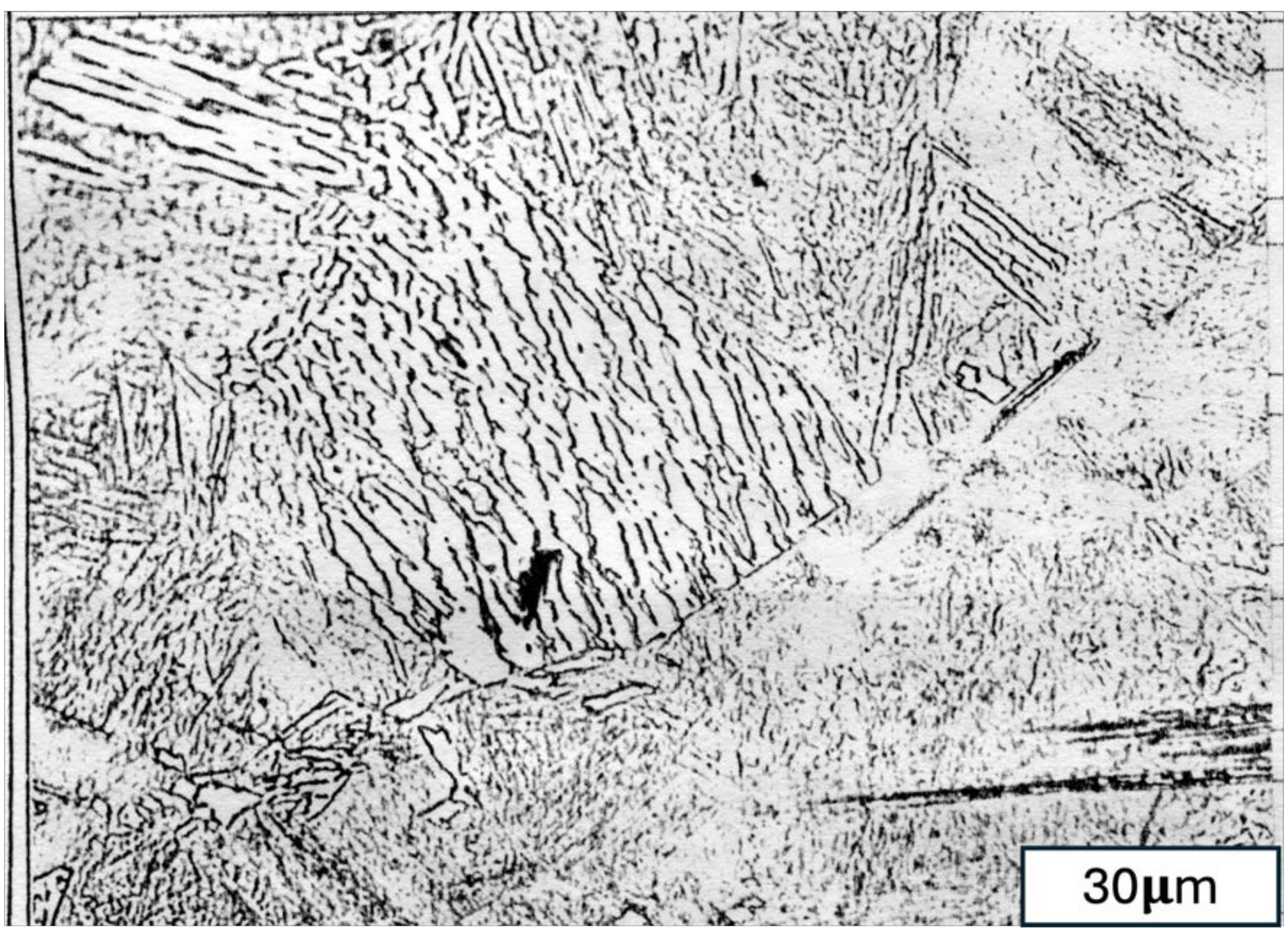

(d)

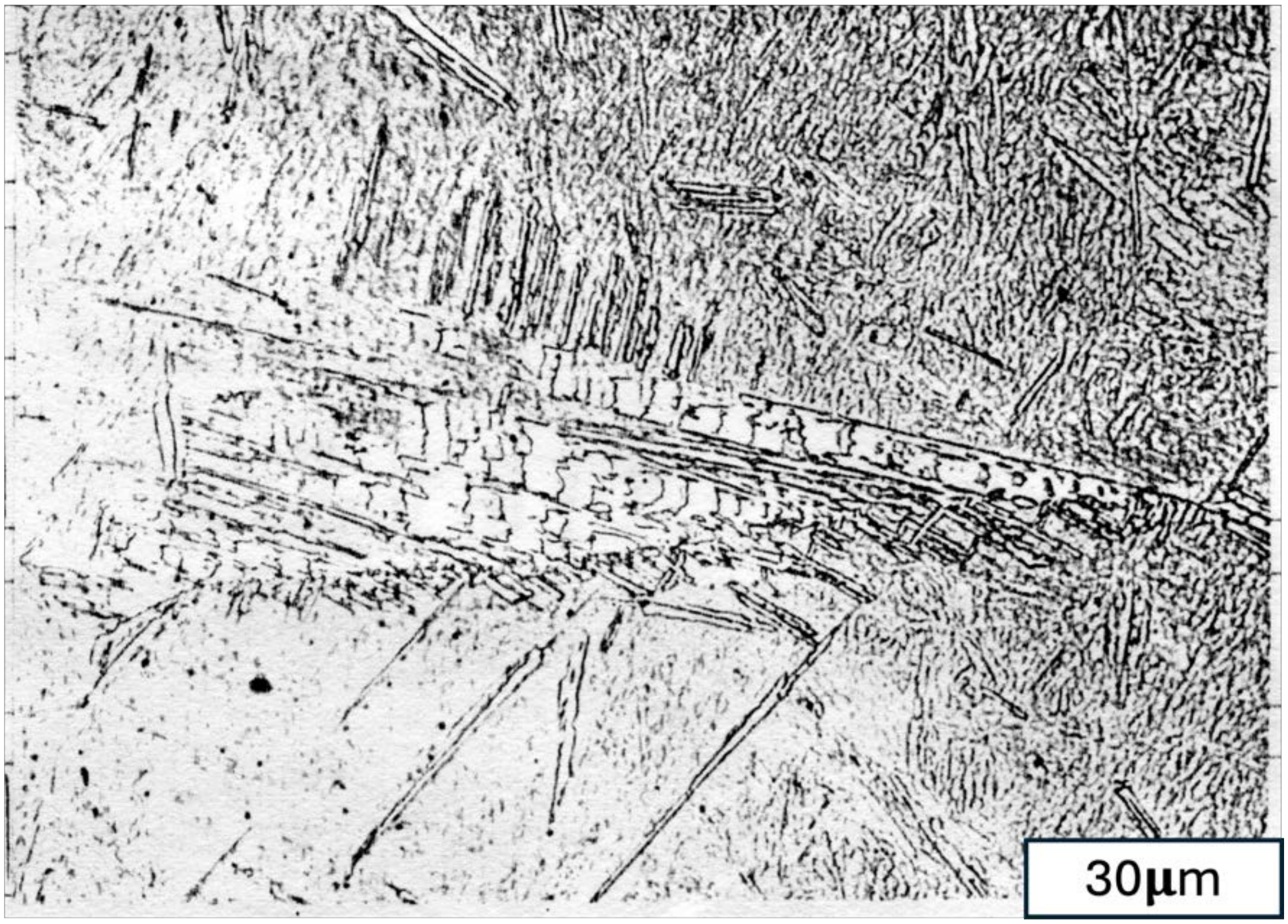

(e)

Figure 5: SEM micrographs of bainite on two-perpendicular surfaces showing the three-dimensional morphologies of ferrite in bainite: (a) a single straight plate in SEM, (b) a single branched plate in SEM, (c) a package with plate shape on upper surface and branched and irregular shapes on lower surface in SEM, (d) a package with 2D plate shape in OM, and (e) a package with 2D rectangle shape. Micrographs were taken from a steel of Fe-0.15C-1.54Mn-0.53Mo-0.015N-0.021RE-0.065V-0.039Al-0.019P-0.015S-0.51Si (wt%) homogenized at 1200°C for one hour and held at 430°C for 5 seconds followed by quenching into water [31].

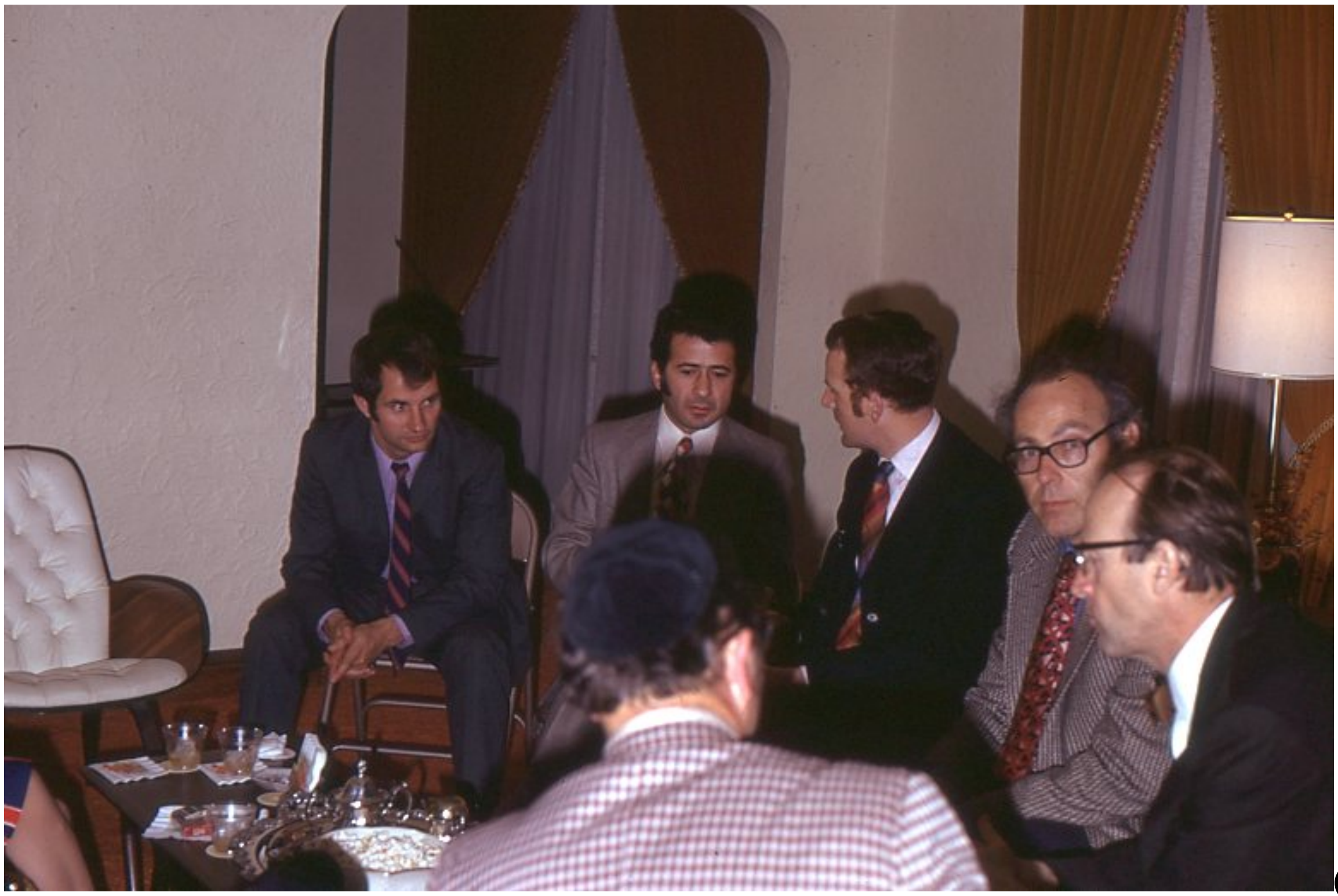
(a)

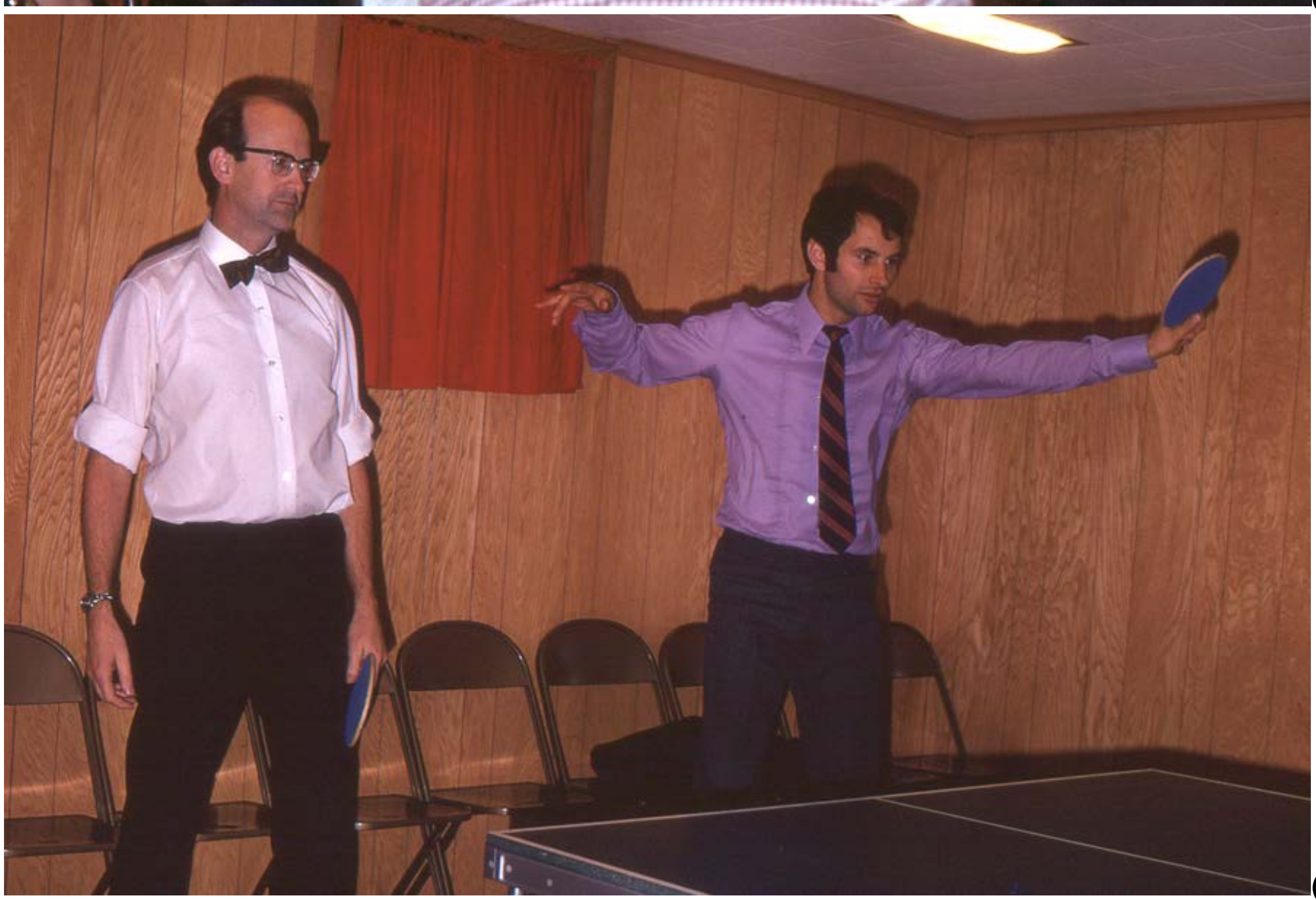
(b)

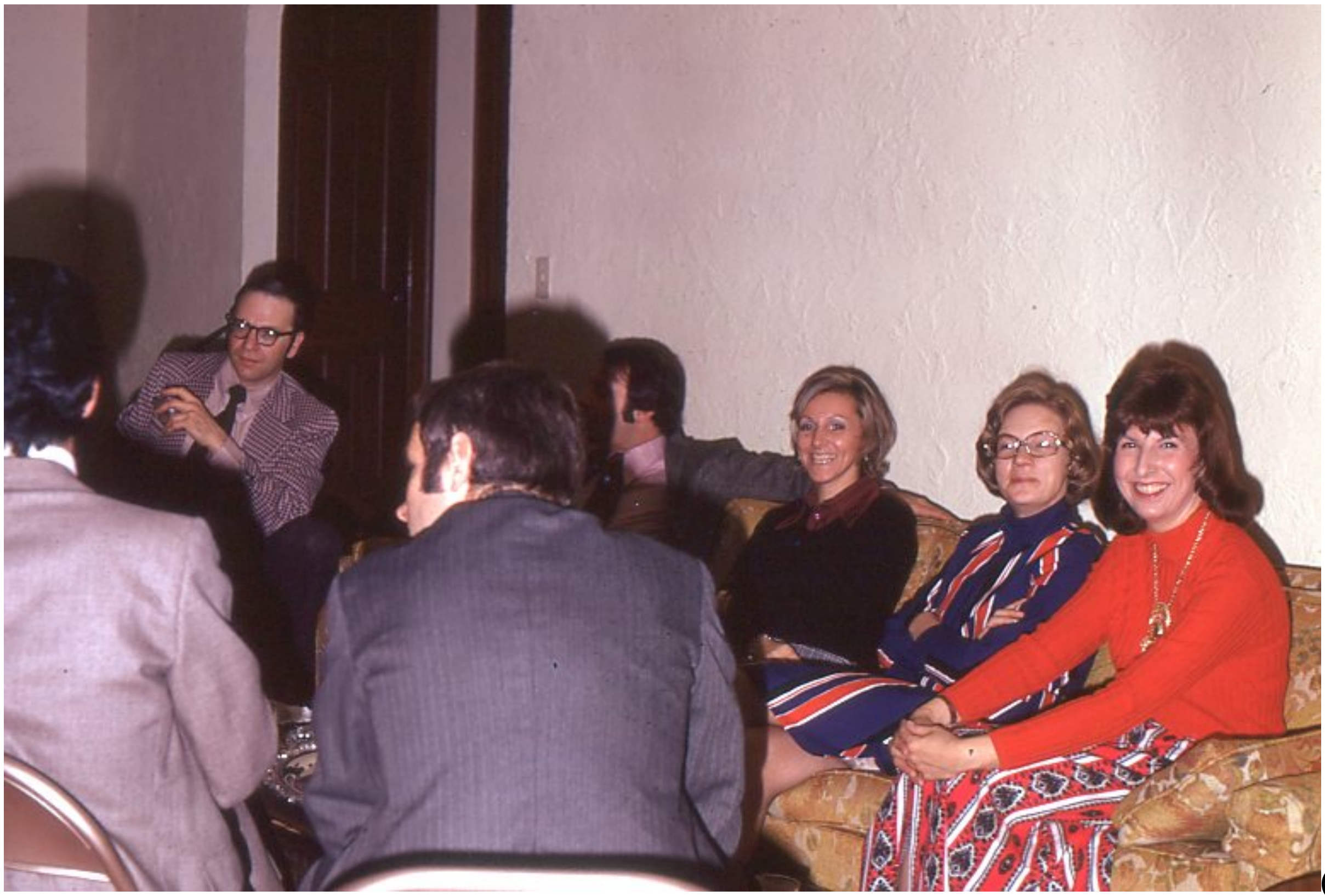(c)

Figure 6: Photos from the first CALPHAD meeting in 1973 in Kaufman's home, (a) Hillert (1*st* from right), Cahn (2*nd* from right), Spencer (3*rd* from right), and Kaufman (showing back of his head); (b) Hillert (left) playing table tennis; (c) Kaufman (facing the camera) and his wife (1st from right).

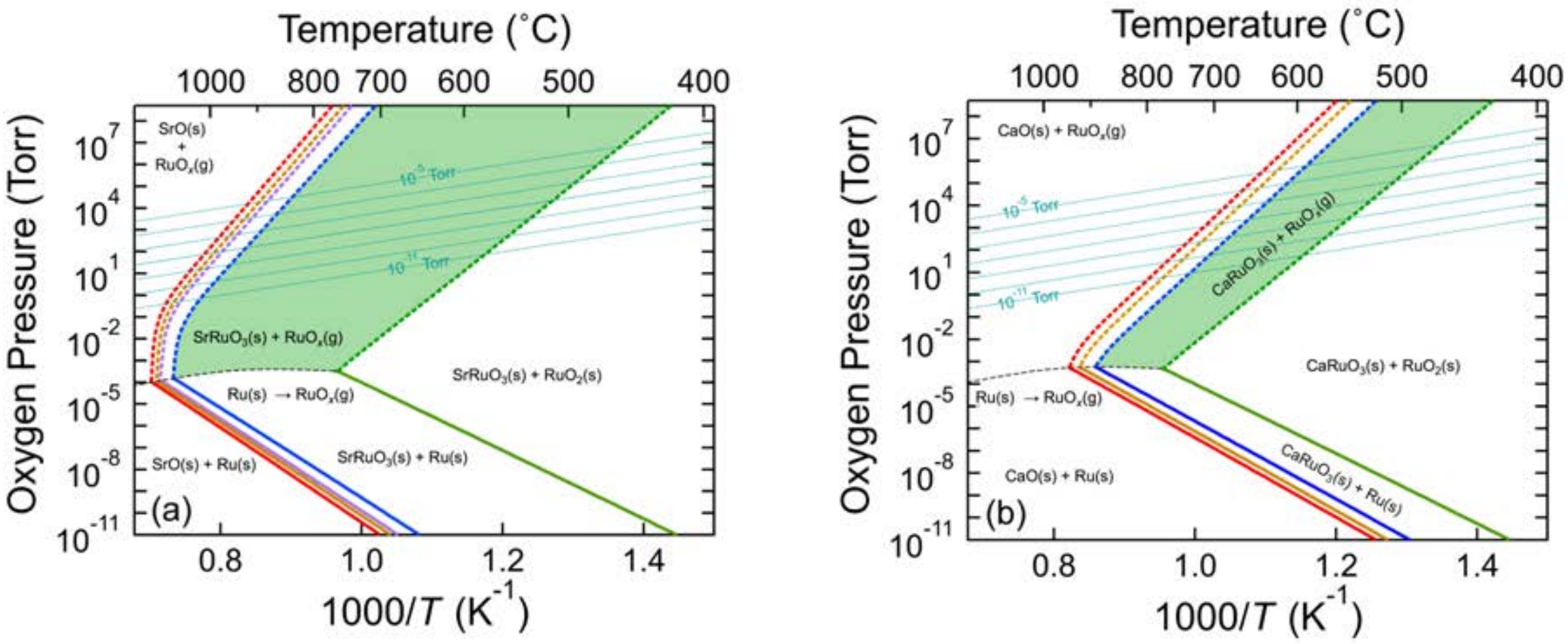

(a) (b)

Figure 7: Thermodynamics of MBE (TOMBE) illustrating the adsorption-controlled growth window for (a) $Sr_{n+1}Ru_nO_{3n+1}$ and (b) $Ca_{n+1}Ru_nO_{3n+1}$ phases. The light green shaded regions in (a) and (b) are the adsorption-controlled growth windows for $SrRuO_3$ and $CaRuO_3$, respectively [93]. The cyan lines show the equivalent oxidation potential for ozone partial pressures ranging from $10^{-11}$ to $10^{-5}$ Torr (1 Torr=133Pa). An excess ruthenium flux of $1.95x10^{17}$ atoms/$m^2$s and $3.2x10^{17}$ atoms/$m^2$s was used in the thermodynamic calculations for (a) and (b), respectively. Reproduced under a Creative Commons Attribution (CC BY) license.